\documentclass[10pt,a4paper,twocolumn,accepted=2026-07-06]{quantumarticle} % format
\pdfoutput=1
\usepackage[british]{babel}
\usepackage{microtype}

\usepackage[sort&compress,numbers]{natbib}
\makeatletter
\AtBeginDocument{%
  \let\origBibitemShut\BibitemShut
  \def\BibitemShut#1{\unskip\origBibitemShut{#1}}%
}
\makeatother

\usepackage[normalem]{ulem}

\usepackage{amssymb}
\usepackage{amsthm}
\newtheorem*{externaltheorem}{Theorem}
\usepackage{braket}
\NewDocumentCommand{\ketbra}{mm}{\ket{#1}\mkern-4mu{\bra{#2}}}
\usepackage{mathtools}
\usepackage{bbm}

\DeclarePairedDelimiter{\abs}{\lvert}{\rvert}
\DeclareMathOperator{\Ei}{Ei}
\DeclareMathOperator{\Tr}{Tr}
\DeclareMathOperator{\tr}{tr}
\newcommand{\id}{\mathbb{I}}
\NewDocumentCommand{\params}{}{\ensuremath{\boldsymbol{\theta}}}

\let\Re\relax
\let\Im\relax
\DeclareMathOperator{\Re}{Re}
\DeclareMathOperator{\Im}{Im}

\usepackage[dvipsnames]{xcolor} % colours
\definecolor{myblue}{RGB}{34,31,150}
\definecolor{mygreen}{RGB}{34,150,31}

\usepackage[breaklinks=true,colorlinks=true,linkcolor=myblue,urlcolor=myblue,citecolor=myblue]{hyperref} % hyperlinks

\begin{document} % manuscript

\title{Measurement incompatibility in Bayesian multiparameter quantum estimation}

\author{Francesco Albarelli}
\orcid{0000-0001-5775-168X}
\email{francesco.albarelli@gmail.com}
\affiliation{Università di Parma, Dipartimento di Scienze Matematiche, Fisiche e Informatiche, I-43124 Parma, Italy}
\affiliation{INFN—Sezione di Milano-Bicocca, gruppo collegato di Parma, I-43124 Parma, Italy}

\author{Dominic Branford}
\orcid{0000-0001-8989-7464}
\email{dominic.branford@unifi.it}
\affiliation{Dipartimento di Fisica e Astronomia, Università di Firenze, via G. Sansone 1, I-50019  Sesto Fiorentino (FI), Italy}

\author{Jes\'{u}s Rubio}
\orcid{0000-0002-8193-8273}
\email{jesus@rubiojimenez.com}
\affiliation{School of Mathematics and Physics, University of Surrey, Guildford GU2 7XH, United Kingdom}
\affiliation{Department of Physics and Astronomy, University of Exeter, Stocker Road, Exeter EX4 4QL, United Kingdom}

\date{3rd August 2026}

\begin{abstract}
We present a comprehensive and pedagogical formulation of Bayesian multiparameter quantum estimation.
Within this framework, we analyse the role of measurement incompatibility and establish its quantitative effect on attainable precision. 
We achieve this by deriving upper bounds based on the pretty good measurement---a notion from hypothesis testing---combined with the evaluation of the Nagaoka--Hayashi lower bound. 
In general, we prove that, as in the many-copy regime of local estimation theory, incompatibility can at most double the minimum loss relative to the idealised scenario in which individually optimal measurements are assumed jointly implementable. 
Therefore, in practical situations, the latter may provide a sufficient and computationally efficient benchmark without solving the full optimisation problem. 
Our results, which we illustrate through applications of discrete phase imaging, phase and dephasing estimation, and qubit sensing, provide analytical and numerical tools for assessing ultimate precision limits and the role of measurement incompatibility in Bayesian multiparameter quantum metrology, including an open-source package for all the bounds discussed here.
\end{abstract}

\maketitle

\section{Introduction}

Quantum metrology explores the fundamental and practical impact of quantum mechanical phenomena, such as superposition, squeezing, and entanglement, on high-precision measurements of physical parameters~\cite{giovannetti2011advances,Polino2020,Liu2022c,Jiao2023a,Montenegro2025b}. 
Its central aim is to surpass the precision limits of classical strategies that do not rely on these quantum effects.
Even if such advantage is not always feasible, a proper understanding of the fundamental precision limits imposed by quantum mechanics is crucial to develop new technologies and guide resource allocation~\cite{demkowiczdobrzanski2015quantum, sahota2015quantum, friis2017flexible}.

Quantum estimation theory (QET) provides the mathematical framework for analysing the precision limits achievable with quantum systems. 
Its most celebrated result is the quantum Cramér--Rao bound (CRB)~\cite{helstrom1976quantuum}.
While many results in quantum metrology rely on the CRB, this approach has important limitations.
Roughly speaking, the CRB is a meaningful figure of merit only when a large number of experiment repetitions is performed, with a significant caveat that it is often hard to understand a priori how many repetitions are needed.
This flavour of QET is known as \emph{local}, meaning that its predictions are generally valid when the discrepancy between the estimates and the true value of the parameters is sufficiently small, which \emph{eventually} happens as samples accumulate.

Global estimation theory \cite{demkowicz2020multi, gorecki2022multiparameter} represents a complementary approach that is especially relevant when resources are constrained and only a limited number of experimental trials can be performed~\cite{tsang2012zivzakai, rubio2019quantum, valeri2020experimental}.
While several global approaches have been proposed~\cite{Gessner2023,Maccone2017,Gebhart2024,Hervas2025,Hervas2025a,Meyer2025,Mukhopadhyay2025a,Polloreno2023,Allen2025}, the most established and extensively developed framework for going beyond the local estimation paradigm is Bayesian quantum estimation. 
Dating back to the early days of the field~\cite{personick1969thesis,personick1971application,Holevo1973,Holevo1973b,yuen1973multipleparameter,helstrom1974noncommuting,helstrom1976quantuum}, it is nowadays being increasingly applied to a wide range of multiparameter sensing experiments involving both photons~\cite{DAurelio2022a,Valeri2023,Cimini2024,Cimini2024a,Belliardo2024b,Minati2025}, atoms~\cite{glatthard2022optimal,overton2024five}, and solid-state systems \cite{espinos2025microwavefree}.
This is the framework adopted in the present work.

Bayesian estimation is not only practically significant in regimes with limited experimental data but also holds deep fundamental implications.
For example, it clarifies impractical claims arising from applying local QET beyond its valid domain, such as the alleged effectiveness of super-Heisenberg-limited schemes~\cite{rivas2012subheisenberg, tsang2012zivzakai, pezze2013subheisenberg} or the apparent enhancement in multiple phase estimation~\cite{humphreys2013quantum, knott2016local, rubio2020bayesian,Gorecki2022}, and it reveals the need to correct the Heisenberg limit in phase estimation by a factor of $\pi$~\cite{gorecki2020pi}. 
In general, the strategies predicted by local QET can become suboptimal in the non-asymptotic regime~\cite{Meyer2025}, where the Bayesian framework consistently enables efficient information extraction~\cite{rubio2019quantum, ravell2025knowledge}.
This is unsurprising given that Bayesian inference constitutes the most basic form of \emph{learning}~\cite{jaynes2003probability}, which is the most comprehensive approach for characterising quantum systems~\cite{Gebhart2023a,Berquin2023,Banchi2024a,Chen2024h,Oh2024,Dutt2024,Wang2025c,Hu2025a,Belliardo2025}.

One of the fundamental aspects of quantum theory is the existence of incompatible measurements that cannot be implemented jointly~\cite{Heinosaari2016,Guhne2023}.
In the context of quantum metrology, this phenomenon becomes crucial when multiple parameters must be estimated simultaneously: measurement incompatibility can impose fundamental trade-offs on the achievable precision of different parameters~\cite{Szczykulska2016,Li2024x,Hu2025f,Kull2020}, which have been quantified in various ways~\cite{Razavian2020,Belliardo2021,Candeloro2021c,Candeloro2024}.
Besides this concrete consequence, incompatibility makes multiparameter theory mathematically challenging \cite{demkowicz2020multi}.
While the fundamental precision limits in single-parameter quantum metrology are relatively well-understood, fundamental multiparameter limits remain a subject of active research~\cite{Albarelli2019c,Liu2019d,Pezze2025}.

It is instructive to review the impact of such incompatibility effects on the development of local QET, before drawing a parallel with the Bayesian approach.
A multiparameter version of the quantum CRB was first derived by Helstrom at the dawn of the field, based on the notion of the symmetric logarithmic derivative (SLD)~\cite{Helstrom1967,helstrom1976quantuum}.
We shall call this bound the Helstrom (SLD) CRB.
While this bound proved attainable for single-parameter estimation~\cite{Braunstein1994,Barndorff-Nielsen2000}, it does not account for the effects of incompatibility and is not generally attainable for multiparameter estimation.
In the following decades, further quantum generalisations of the CRB were developed, providing a better account of measurement incompatibility, and culminating in the discovery that the ultimate precision limit attainable with collective (entangling) measurements over many copies is given by the Holevo–Cramér–Rao bound (HCRB)~\cite{demkowicz2020multi,Hayashi2008a,Kahn2009,Yamagata2013,Yang2018a,Fujiwara2023}.
It was eventually understood that, in this asymptotic scenario where many copies can be measured collectively, the effect of measurement incompatibility is relatively mild: the HCRB can be at most twice the original bound obtained by Helstrom~\cite{Tsang2019,Carollo2019}
(see also Refs.~\cite{Albarelli2019c,demkowicz2020multi,Belliardo2021}).

By contrast, the fundamental precision achievable with separable measurements on each copy of the state is not efficiently computable in local QET, even numerically (although collective measurements are unnecessary for pure states~\cite{Matsumoto2002}).
Nonetheless, several bounds for this setting have been derived~\cite{Gill2000,Lu2020a,Chen2022f,Chen2022g,Chen2024j,Wang2025f}, among which we highlight the Nagaoka–Hayashi (NH) bound~\cite{Nagaoka1989,Nagaoka1991,hayashi1999simultaneous,Conlon2020,Conlon2024c,Yung2024,Das2024,Conlon2025a}.
Most recently, the problem has been studied using conic programming~\cite{Hayashi2023a}, providing a unifying framework and new tools that have already been applied to probe state optimisation~\cite{Hayashi2024}.

Despite significant advances in local QET, the role of measurement incompatibility in Bayesian estimation has received much less attention.
The general conditions that a Bayesian estimation strategy must satisfy to be optimal were already identified in the aforementioned early works~\cite{Holevo1973,Holevo1973b,helstrom1974noncommuting,helstrom1976quantuum}, including a discrete analogue for hypothesis testing~\cite{Yuen1970,Yuen1975}. 
A generalisation to convex functions has also been proposed~\cite{Coutts2018}.
Ref.~\cite{helstrom1976quantuum} provided examples of how these conditions can be used to verify whether a proposed strategy is indeed optimal. 
Reversing this procedure---namely, using the conditions to derive the optimal strategy---is, however, exceedingly challenging, with only a limited number of problems where this can be done analytically. 
For instance, this procedure can be carried out in the simple case of two parameters~\cite{brody1996bayesian}.
 
For single-parameter estimation, the class of lo\-ca\-tion-iso\-mor\-phic estimation problems can be optimised exactly~\cite{rubio2024first}, and the solution can be shown to satisfy the corresponding optimality conditions. 
This family includes, as a particular case, the optimisation of the standard mean square error---called mean square loss (MSL) in this work---a problem originally solved by Personick~\cite{personick1971application}. 
While Personick made an early attempt to extend this reasoning to the estimation of two parameters~\cite{personick1969thesis}, closed-form solutions based on indirect measurements were subsequently obtained~\cite{Belavkin1973}, and a solution for the estimation of two displacement parameters in Gaussian states with Gaussian priors is known~\cite{wang2007}, a general and practical procedure for determining optimal solutions in the Bayesian estimation of multiple parameters has remained elusive for decades.

Only recently has a multiparameter bound, analogous to the Helstrom (SLD) CRB---and thus neglecting incompatibility effects---been derived~\cite{rubio2020bayesian}.
As motivated later, we propose referring to this as the symmetric posterior mean (SPM) bound.
Bayesian analogues of several other local bounds, including the NH bound, have since been established, incorporating measurement incompatibility to a certain extent~\cite{Suzuki2024a,Zhang2024g}.  
In addition, numerical algorithms for identifying optimal estimation strategies are now under active investigation~\cite{Kaubruegger2023,Bavaresco2024designing}.

Continuing the parallel between the local and Bayesian frameworks, this work addresses a key question: how strongly does measurement incompatibility affect Bayesian multiparameter QET?
In other words, what is the maximal discrepancy between the achievable precision and the bound that neglect incompatibility?

\textit{Contributions.}---We show that, in analogy with local QET, the minimum MSL can be at most twice the SPM benchmark that is obtained when measurement incompatibility is ignored. 
Key differences, however, are that this upper bound now holds at the single-shot level, but can also be trivial, i.e., larger than the MSL derived solely from \emph{a priori} information. 
This effect stems from the central role of prior information in Bayesian estimation, which, as illustrated through examples in quantum imaging, phase and dephasing estimation, and qubit sensing, may significantly modify the apparent incompatibility of the estimation protocol.
Consequently, measurement incompatibility in Bayesian multiparameter QET is not only a relatively mild factor in the overall estimation precision, but it can also be concealed when the influence of the prior is strong.
We further derive a tighter upper bound that is guaranteed to be non-trivial, by combining measurements introduced in hypothesis testing~\cite{Belavkin1975a,Belavkin1975,Kholevo1979,Hausladen1994} with optimal classical Bayes estimators~\cite{jaynes2003probability}.
Our upper bounds are complementary to existing lower bounds tighter than the SPM one, where the Bayesian NH bound~\cite{Suzuki2024a} represents the state of the art.
In addition, we derive a class of lower bounds based on generalised posterior mean operators beyond the symmetric version that, while less tight than the NH bound, may be of independent interest, since they are obtained via the theory of monotone metrics, analogously to established results in local QET~\cite{Hayashi2017c}.
An open-source code for the numerical evaluation of all relevant bounds is available on GitHub~\cite{multipar2025code}.
Finally, beyond deriving bounds on the minimum MSL, we determine explicit conditions for the measurement strategies that achieve it. 
Combined with the pedagogical and self-contained introduction to Bayesian multiparameter quantum metrology presented in this work, these results establish a solid foundation for further advances in the field.

\section{Bayesian multiparameter quantum estimation theory}

\subsection{Preliminaries}

Let $\boldsymbol{\Theta}$ denote a vector of unknown parameters and $\boldsymbol{\theta} \in \Omega$ a vector hypotheses---values the unknown parameter could take---both understood as $d{\times}1$ column vectors in $\mathbb{R}^d$.
Here, $\Omega$ represents a generic parameter space. 
The information about~$\boldsymbol{\Theta}$ available prior to any measurement is encoded in a probability density $p(\boldsymbol{\theta})$, satisfying the non-negativity and normalisation conditions:
\begin{equation}
    p(\boldsymbol{\theta}) \ge 0, \quad 
    \int_{\Omega} \mathrm{d}\boldsymbol{\theta}\, p(\boldsymbol{\theta}) = 1.
\end{equation}
This \( p(\params) \) is referred to as the \emph{prior probability}.

To gather information about $\boldsymbol{\Theta}$ via measurements, the vector hypothesis is encoded in the quantum state of a suitable physical system, such as an atomic ensemble or a Mach--Zehnder interferometer, denoted by $\rho(\boldsymbol{\theta})$. 
This is a positive semidefinite operator with unit trace:
\begin{equation}
\rho(\boldsymbol{\theta}) \succeq 0, \quad 
\operatorname{Tr}[\rho(\boldsymbol{\theta})] = 1.
\end{equation}
We focus on finite-dimensional quantum systems, and whilst we expect most results in Sec.~\ref{sec:incompatibility} to hold more generally, we leave a more rigorous analysis for future work.

Once a state $\rho(\boldsymbol{\theta})$ is prepared, a measurement with outcome $x \in \chi$ is performed, where $\chi$ denotes the sampling space. 
Each outcome is associated with an element $M(x)$ of a positive operator-valued measure (POVM), which is positive semidefinite and resolves the identity:
\begin{equation}
    M(x) \succeq 0, \quad \int_{\chi} \mathrm{d}x\, M(x) = \mathbb{I}.
\end{equation}
For simplicity, in Sec.~\ref{sec:applications}, we only consider examples with discrete outcomes, which is natural when focusing on finite-dimensional systems. 
In such cases, continuous outcomes can be realised by performing finite measurements depending on a classical parameter~\cite{chiribella2007how}.
Conversely, discrete outcomes can be embedded in a continuous-outcome formulation by expressing $M(x)$ as a sum of Dirac delta functions centered on the discrete outcome $n$,
\begin{equation}
        \label{eq:discrete-POVM}
       M(x) = \sum_n  M_n\, \delta(x - n),
\end{equation}
associated with a POVM element $M_n$ satisfying $M_n \succeq 0$ and $\sum_n M_n = \mathbb{I}$.
The continuous-outcome formalism is particularly convenient for variational calculus, which is frequently employed in Bayesian optimisation \cite{personick1971application, jaynes2003probability}.

At this point, Bayesian inference offers a straightforward solution to the problem of inferring the value of $\boldsymbol{\Theta}$ by combining Bayes' theorem with the Born rule. 
Specifically, the prior probability $p(\boldsymbol{\theta})$ is \emph{updated} according to
\begin{equation}
    p(\boldsymbol{\theta}) \mapsto p(\boldsymbol{\theta}|x) \propto p(\boldsymbol{\theta}) \, \Tr[M(x) \rho(\boldsymbol{\theta})].
\end{equation}
Here, the probability density $p(\boldsymbol{\theta}|x)$ denotes the \emph{posterior probability} while $\Tr[M(x) \rho(\boldsymbol{\theta})] = p(x | \boldsymbol{\theta})$ is identified as the \emph{likelihood function}. 

The posterior probability represents the full state of knowledge about the values of the parameters $\boldsymbol{\theta}$ available after observing the measurement outcome $x$.
In that sense, it could be already considered a complete solution of the problem. 

Calculation of the posterior probabilities underpins all Bayesian measurement protocols. 
However, more ergonomical theoretical constructions are generally required. 
For instance, it is customary to identify a single vector value representing our best guess for the unknown $\boldsymbol{\Theta}$.
Intuitively, one possibility is taking the vector of parameters that maximises the posterior probability, which is particularly suitable in situations with a unimodal Gaussian profile. 
That is,
\begin{equation}
    \boldsymbol{\tilde{\theta}}_{\mathrm{MAP}}(x) = \arg\max_{\boldsymbol{\theta}} p(\boldsymbol{\theta}|x).
\end{equation}
This is the so-called maximum \emph{a posteriori} (MAP) estimator~\cite{linden2014bayesian}, which plays a role similar to that of the maximum likelihood estimator in local QET~\cite{kay1993fundamentals}.

Furthermore, to select a state $\rho(\boldsymbol{\theta})$ and a POVM element $M(x)$ and guide
optimal experimental design, together with an estimator $\boldsymbol{\tilde{\theta}}(x)$ for post-processing the data, one requires a figure of merit that quantifies estimation performance and serves as the optimisation criterion.
To that end, one typically introduces a loss function\footnote{
This function is also often referred to as a cost, deviation, or error function. 
The term \emph{loss} is preferred because it conveys a general notion of estimation inaccuracy, encompassing statistical, systematic, and context-dependent sources.
In contrast, \emph{deviation} and \emph{error} typically emphasise purely statistical discrepancies, whereas \emph{cost} may suggest application-specific considerations more typical of engineering contexts.}
$l(\boldsymbol{\theta}, \boldsymbol{\tilde{\theta}})$ quantifying the inaccuracy that may arise by choosing an estimate $\boldsymbol{\tilde{\theta}}$ should the hypothesis $\boldsymbol{\theta}$ correspond to the true parameter values \cite{jaynes2003probability}.
An estimator, state, and POVM are then said to be optimal when they minimise the mean loss
\begin{equation}
    \mathcal{L} 
    \coloneq \int \mathrm{d}\params\, \mathrm{d}x\, p(\params)\, \Tr[M(x) \rho(\boldsymbol{\theta})]\, l(\boldsymbol{\tilde{\theta}}(x), \params),
    \label{eq:msl}
\end{equation}
where, for simplicity, we omit explicit reference to the domains $\Omega$ and $\chi$. 
This convention will be followed henceforth.

This work evaluates measurement incompatibility within the Bayesian paradigm by establishing lower and upper bounds on a particular type of mean loss $\mathcal{L}$. 
The remainder of this section addresses formal aspects of $\mathcal{L}$, as well as the notion of metrological optimality based on this figure of merit, that are necessary to derive and contextualise our main results in Sec.~\ref{sec:incompatibility}.

\subsection{Estimation performance}
\label{subsec:estimation-performance}

We first observe that Eq.~\eqref{eq:msl} quantifies estimation performance by integrating over all hypothesis values in $\Omega$, whose contribution can vary from small to large, weighted by the prior probability $p(\boldsymbol{\theta})$.
This is the key feature that makes the framework not only global but also Bayesian.
There exists a more conservative, non-Bayesian approach to global estimation: the minimax approach, which is based on the \emph{frequentist} loss~\cite{demkowiczdobrzanski2015quantum}
\begin{equation}
    \mathcal{L} (\boldsymbol{\theta})
    \coloneq \int \mathrm{d}x\, \Tr[M(x) \rho(\boldsymbol{\theta})]\,
      l(\boldsymbol{\tilde{\theta}}(x), \boldsymbol{\theta}),
      \label{eq:local-msl}
\end{equation}
and seeks to minimise the worst-case value of $\mathcal{L}(\boldsymbol{\theta})$
over relevant regions of the parameter space \cite{gorecki2023heisenberg}.
However, since the minimax criterion can be written as the supremum of Eq.~\eqref{eq:msl} over all priors, our bounds in Sec.~\ref{sec:incompatibility} are also relevant for minimax estimation.
Note that, at the other extreme, Eq.~\eqref{eq:local-msl} is employed in local QET by evaluating it in an infinitesimal neighbourhood of specific values of $\params$~\cite{kay1993fundamentals, demkowiczdobrzanski2015quantum}.

We further comment that, in Bayesian estimation, the additional integral over $x$ is required only when searching for optimal estimation protocols, as in our case, since this task precedes the recording of specific outcomes.
For data post-processing, estimation performance must instead be assessed using an outcome-dependent quantifier \cite{jaynes1976confidence}, e.g., 
\begin{equation}
    \mathcal{L}(x) 
    \coloneq \int \mathrm{d}\params\, p(\params | x)\, l(\boldsymbol{\tilde{\theta}}(x), \params).
\end{equation}
Alternatively, one may employ credibility intervals, also based on the posterior \cite{jaynes1976confidence, boeyens2021uninformed}. 
Both experimental error assessment and the minimax and local estimation approaches discussed above are beyond the scope of this work, and we therefore focus exclusively on Eq.~\eqref{eq:msl}.

So far we have not specified any loss function $l$.
The most prevalent choice is
\begin{equation}
    l(\boldsymbol{\tilde{\theta}}, \boldsymbol{\theta}) = (\boldsymbol{\tilde{\theta}} - \boldsymbol{\theta})^T L\,(\boldsymbol{\tilde{\theta}} - \boldsymbol{\theta}),
    \label{eq:square_loss}
\end{equation}
where $L$ is a positive semidefinite weight matrix, i.e., $L \succeq 0$, and we assume it to be parameter-independent. 
The resulting~\(\mathcal{L}\) is the standard MSL, and this loss quantifier leads to the posterior mean (PM) as the optimal Bayes estimator~\cite{jaynes2003probability}:
\begin{equation}
\label{eq:PMestimator}
    \tilde{\theta}_{\mathrm{opt},i}(x) = \int \! \mathrm{d}\params \, p( \params | x)\, \theta_i = \frac{ \int \! \mathrm{d}\params \, p(\params) \, p( x  |  \params ) \, \theta_i}{ \int \! \mathrm{d} \params \, p(\params) \, p( x | \params ) } .
\end{equation}

There is a widespread belief that the MSL applies universally, regardless of the specific system, measurement, or type of parameters of interest.  
Strictly speaking, however, the MSL is only appropriate for location or shift parameters. 
This is because the topology for such parameters corresponds to the real line, so that the corresponding loss function must be translation invariant~\cite{jaynes2003probability}, i.e.,
\begin{equation}
    l(\boldsymbol{\tilde{\theta}} + \boldsymbol{\gamma}, \boldsymbol{\theta} + \boldsymbol{\gamma}) = l(\boldsymbol{\tilde{\theta}}, \boldsymbol{\theta}),
\end{equation}
as Eq.~\eqref{eq:square_loss} is\footnote{While one can generally consider parameter-dependent weight matrices, these break the translation invariance of Eq.~\eqref{eq:square_loss}.}.
Here, $\boldsymbol{\gamma}$ is an arbitrary constant vector.

In contrast, parameter types with different properties require alternative loss functions.  
Cycle parameters, for instance, are estimated using trigonometric loss functions---phases in interferometry are the archetypical example~\cite{helstrom1976quantuum,holevo2011probabilistic, demkowiczdobrzanski2015quantum}. 
Moreover, recent results show that scale parameters such as temperature or decay rate require logarithmic loss functions for consistent estimation~\cite{rubio2021nov}, while weight parameters such as probability of success or coherence parameters demand hyperbolic loss functions~\cite{rubio2025on}.
Naive use of the MSL, while it may occasionally yield correct results---particularly when $\boldsymbol{\Theta}$ is known to lie close to a reference point~$\boldsymbol{\Theta}_0$---is physically unjustified and therefore discouraged unless there is no better choice.

Despite this, one can retain the mathematical simplicity of the MSL while respecting parameters' more general topologies in certain scenarios.  
Specifically, Refs.~\cite{rubio2024first, tsang2022generalized} demonstrated that many parameter types admit mean losses equivalent to the MSL, up to a mapping that, in the context of multiparameter estimation, can be expressed as
\begin{equation}
    l(\boldsymbol{\tilde{\theta}}, \boldsymbol{\theta}) 
    = (\boldsymbol{f}(\boldsymbol{\tilde{\theta}}) - \boldsymbol{f}(\boldsymbol{\theta}))^T 
    L\,
    (\boldsymbol{f}(\boldsymbol{\tilde{\theta}}) - \boldsymbol{f}(\boldsymbol{\theta})),
    \label{eq:symmetry-msl}
\end{equation}
where $\boldsymbol{f}$ is chosen according to symmetries of the problem~\cite{rubio2025on}; see Ref.~\cite{overton2024five} for an experimental demonstration with cold $\mathrm{Cs}$ atoms. 
Note that phases fall outside this framework~\cite{holevo2011probabilistic}, though they can be approximated by the MSL itself when confined to intervals of order $\pi/2$~\cite{friis2017flexible, rubio2019quantum}.

Our results in Sec.~\ref{sec:incompatibility} are formulated in terms of the standard MSL for pedagogical reasons, but the extension to Eq.~\eqref{eq:symmetry-msl} is immediate.  
While a general multiparameter symmetry-informed treatment is deferred, Sec.~\ref{sec:phase-dephasing} provides an illustration of how to use Eq.~\eqref{eq:symmetry-msl}.  

\subsection{Optimal quantum strategies}\label{sec:opt-quantum-strategies}

We proceed by providing the conditions that minimise the MSL and tools to establish an operational notion of measurement incompatibility in Bayesian multiparameter quantum estimation.
For a fixed measurement, the optimal classical estimator is the posterior mean (PM), already introduced in Eq.~\eqref{eq:PMestimator}.
In the quantum setting, however, one must also minimise the MSL over all generalised measurements, thereby making the optimisation substantially more challenging.

To ease the manipulation of multiparameter expressions, we first rewrite the MSL as
\begin{equation}
    \mathcal{L} = \lambda + L^{ij}\,\Tr[\rho_0 M_{ji} - 2 \bar{\rho}_j M_i].
    \label{eq:msl_compact}
\end{equation}
Here, $\lambda$ is the weighted second moment of the prior probability,  
\begin{equation}
    \lambda \coloneq \int \mathrm{d}\params\, p(\params)\, \boldsymbol{\theta}^T L\, \boldsymbol{\theta},
    \label{eq:prior-second-moment}
\end{equation}
which differs from the \emph{a priori} MSL for nonvanishing prior mean, i.e., when \( \int \mathrm{d}\params\, p(\params)\, \theta_i \neq 0 \) for at least some $i$; see Eq.~\eqref{eq:prior-loss} and App.~\ref{app:prior-loss}. 
The element $L^{ij}$ denotes the $ij$-th component of $L$, and we employ the Einstein summation convention over indices $i,j = 1,\dots,d$.

We further define, respectively, the following zeroth and first state moments: 
\begin{subequations}
\begin{align}\label{eq:state-moments}
    \rho_0 &:= \int \mathrm{d}\params\, p(\params)\, \rho(\params), \\
    \bar{\rho}_i &:= \int \mathrm{d}\params\, p(\params)\, \rho(\params)\, \theta_i ,
    \label{eq:first-state-moment}
\end{align}
\end{subequations}
as well as the following first and second measurement moments:
\begin{subequations}
\label{eq:measurement-moments}
\begin{align}
    M_i &:= \int \mathrm{d}x\, M(x)\,\tilde{\theta}_i(x) = \int \mathrm{d}\boldsymbol{\tilde{\theta}}\, M(\boldsymbol{\tilde{\theta}})\,\tilde{\theta}_i , \\
    M_{ij} &:= \int \mathrm{d}x\, M(x)\,\tilde{\theta}_i(x) \tilde{\theta}_j(x) 
    = \int \mathrm{d}\boldsymbol{\tilde{\theta}}\, M(\boldsymbol{\tilde{\theta}})\,\tilde{\theta}_i \tilde{\theta}_j,
\end{align}
\end{subequations}
where to obtain the rightmost expressions we have combined the estimators and the measurement by using the relabelled POVM elements~\cite{demkowicz2020multi} 
\begin{equation}
    M(\boldsymbol{\tilde{\theta}}) = \int \mathrm{d}x\, M(x)\,\delta(\boldsymbol{\tilde{\theta}} - \boldsymbol{\tilde{\theta}}(x));
    \label{eq:Mtheta}
\end{equation}
this will be useful in following calculations.

Given that $p(\params)$ must be fixed \emph{a priori}, and $L$ is fixed according to the priority assigned to certain linear combinations or specific parameters, $\lambda$ remains constant throughout.
The degrees of freedom for optimising the MSL are thus $\rho(\boldsymbol{\theta})$ and either the pair $(\boldsymbol{\tilde{\theta}}(x), M(x))$ or, as we will consider from now on, $M(\boldsymbol{\tilde{\theta}})$.
Since our focus is measurement incompatibility, and not the trade-offs that arise from the choice of probing strategy before the final measurement~\cite{ragy2016compatibility,Albarelli2022,Hu2024a}, we shall also assume that $\rho(\boldsymbol{\theta})$---and therefore $\rho_0$ and $\bar{\rho}_i$---is fixed.

The general conditions constraining the form of the optimal $M(\boldsymbol{\tilde{\theta}})$, denoted $M_{\mathrm{opt}}(\boldsymbol{\tilde{\theta}})$, have been established since the seminal works of Personick, Helstrom, Holevo, and others \cite{personick1971application, Holevo1973, Holevo1973b, yuen1973multipleparameter, helstrom1976quantuum,kholevo1978investigations} in the 1970s.
In App.~\ref{app:fundamental-eqs}, we revisit these conditions and show that, for the MSL, they read
\begin{equation}
\begin{aligned}
    L^{ij} [&\;(\tilde{\theta}_j \tilde{\theta}_i \mathbb{I} - M_{\mathrm{opt}, ji})\rho_0
    + \rho_0(\tilde{\theta}_j \tilde{\theta}_i \mathbb{I} - M_{\mathrm{opt}, ji}) \\[6pt]
    &- 2 (\tilde{\theta}_j \mathbb{I} - M_{\mathrm{opt}, j})\bar{\rho}_i
    - 2 \bar{\rho}_j(\tilde{\theta}_i \mathbb{I} - M_{\mathrm{opt}, i})] \succeq 0,
\end{aligned}
\label{eq:msl-opt-condition}
\end{equation}
for all $\boldsymbol{\tilde{\theta}}$.
Here, $M_{\mathrm{opt}, ij}$ and $M_{\mathrm{opt}, i}$ are obtained by inserting $M_{\mathrm{opt}}(\boldsymbol{\tilde{\theta}})$ into Eqs.~\eqref{eq:measurement-moments}.
This leads to the minimum mean loss
\begin{equation}
    \mathcal{L}_\mathrm{min} = \mathcal{L}[M_{\mathrm{opt}}(\boldsymbol{\tilde{\theta}})],
    \label{eq:msl-min}
\end{equation}
i.e., $\mathcal{L}$ in Eq.~\eqref{eq:msl_compact} evaluated at the optimal POVM.

Eq.~\eqref{eq:msl-opt-condition} constitutes our first result.  
In general, determining the explicit form of $M_{\mathrm{opt}}(\boldsymbol{\tilde{\theta}})$ from Eq.~\eqref{eq:msl-opt-condition} is a nontrivial task.  
Nevertheless, these conditions provide a practical criterion for verifying whether a candidate POVM is truly optimal.  
Note that employing a symmetry-informed yet still quadratic loss function [Eq.~\eqref{eq:symmetry-msl}] yields optimality conditions of the same structure, with the parameter space remapped according to the vector symmetry function $\boldsymbol{f}$.

\subsection{Lower bounds from optimal single-parameter strategies}

While exact computation of $\mathcal{L}_\mathrm{min}$ is generally difficult, it admits a lower bound~\cite{rubio2020bayesian,Sidhu2019,demkowicz2020multi}
\begin{equation}
    \mathcal{L}_{\mathrm{min}} \geq 
    \lambda - L^{ij} \Tr [\rho_0 S_j S_i] \eqqcolon \mathcal{L}_{\mathrm{SPM}},
    \label{eq:SPM_lower_bound}
\end{equation}
where $\{ S_i \}$ are the SPM operators, defined as the solutions to the Lyapunov equations
\begin{equation}
    \bar{\rho}_i  = \frac{S_i \rho_0 + \rho_0 S_i}{2} 
    = E_{\rho_0,\mathrm{S}} \left[S_i \right].
    \label{eq:unconstrained_condition_constantL}
\end{equation}
For later convenience, we have introduced the anticommutator map (superoperator)
\begin{equation}
E_{\rho,\mathrm{S}}[X] \coloneq \frac{1}{2}\left( X \rho + \rho X \right) ,
\label{eq:E-map-SLD}
\end{equation}
which admits a compact expression ${S_i = E_{\rho_0,\mathrm{S}}^{-1}[ \bar{\rho}_i]}$ for the SPMs, and thus
\begin{equation} 
\mathcal{L}_{\mathrm{SPM}} = \lambda - L^{ij} \Tr [\bar{\rho}_j E_{\rho_0,\mathrm{S}}^{-1}[ \bar{\rho}_i] ].
\end{equation}
Here and in the rest of this work, we assume that this inverse exists, which is guaranteed for a full-rank $\rho_0$.
When this is not the case, one simply needs to work in the relevant subspace spanned by the parametric family $\rho(\params)$, which corresponds to the support of $\rho_0$%
\footnote{This is possible because the support of the operators $\{ \bar{\rho}_j \}$ is contained in the support of $\rho_0$, since $\braket{\psi | \rho_0 | \psi} = 0 $ implies $\braket{\psi | \rho(\params) | \psi} = 0$ almost everywhere (in the measure-theoretic sense), which in turn implies $\braket{\psi | \bar{\rho}_j | \psi} = 0$.
This is markedly different from local QET, where the partial derivatives operators may have components also in the kernel of the state, and properties
of rank-deficient models are non-trivial.
For example, necessary and sufficient conditions for attainability of the SLD CRB are not fully characterised~\cite{Yang2018b,horodeckiFiveOpenProblems2022,Conlon2024c,Nurdin2024a} (except for pure states~\cite{Matsumoto2002,Pezze2017})}.

We note that several names have been adopted in the recent literature for the operators $\{ S_i \}$, mostly in analogy with the SLD operators that appear in local QET.
We propose instead the term SPM operators, by analogy with the optimal classical Bayes estimator for each parameter in Eq.~\eqref{eq:PMestimator}.

Formally, the PM estimator is the component-wise ratio of the first-moment probability vector \(\int \mathrm{d}\params\, p(\params)\, p(x|\params) \, \theta_i\) to the average probability vector \(\int \mathrm{d} \params\, p(\params)\, p(x|\params)\).
Intuitively, if \(p(x|\params)\) is formally replaced by a density operator \(\rho(\params)\), there are infinitely many inequivalent ways to ``divide'' $\bar{\rho}_i$ by $\rho_0$. The SPMs correspond to the symmetric choice, where ``multiplication by $\rho_0$'' is implemented by the map $E_{\rho_0,\mathrm{S}}$.
Infinitely many other choices are possible and become inequivalent whenever $\bar{\rho}_i$ and $\rho_0$ do not commute, leading to different lower bounds; see App.~\ref{app:matrix-lower-bounds}.
This intuitive reasoning was originally identified by Personick \cite{personick1971application} and is semantically consistent with local QET, in which the logarithmic derivative of a probability distribution admits inequivalent operator generalisations in the quantum setting~\cite{Hayashi2017c}.

Crucially, the SPM bound has the additional property of being attainable in single-parameter estimation by projecting onto the eigenbasis of the single SPM operator available in that case~\cite{personick1971application,rubio2020bayesian,demkowicz2020multi,rubio2023quantum}.
Consequently, upon performing the eigendecomposition
\begin{equation}
    S_i = \sum_n \ketbra{s_{i,n}}{s_{i,n}} s_{i,n}
\end{equation}
for the $i$-th operator, where we assume a discrete eigenbasis for simplicity but without loss of generality,  
the spectrum $\{ s_{i,n} \}$ defines the estimator (see Refs.~\cite{personick1971application, rubio2023quantum} for details, including a continuous-variable formulation)
\begin{equation}
    s_{i,n} = \int \mathrm{d}\theta\, p(\theta|s_{i,n})\,\theta 
    = \tilde{\theta}_{\mathrm{opt}, i}(n),
    \label{eq:classical_PM_estimator}
\end{equation}
which naturally corresponds to the posterior mean (PM) estimator for the $i$-th parameter.

While the PM estimators remain optimal in the multiparameter case~\cite{rubio2020bayesian}, the SPM bound is generally not attainable due to the non-commutativity of the SPM operators. 
Nevertheless, it provides a valid lower bound on the minimum MSL and can be interpreted as the optimum obtained when incompatibility among the individually optimal measurements is neglected.
Accordingly, the next section introduces a notion of measurement incompatibility based on $\mathcal{L}_{\mathrm{SPM}}$, establishing it as a result of Bayesian estimation theory analogous in fundamental significance to the Helstrom (SLD) CRB in local QET~\cite{demkowicz2020multi}.

\section{Measurement incompatibility: definition and bounds}\label{sec:incompatibility}

\subsection{Incompatibility figure of merit}

As we have just argued, the benchmark to evaluate the impact of measurement incompatibility is the lower bound on the MSL, $\mathcal{L}_{\mathrm{SPM}}$, which is achievable when all SPM operators share a common eigenbasis and thence are compatible observables.
Therefore, measurement incompatibility in Bayesian multiparameter estimation can be quantified by comparing $\mathcal{L}_{\mathrm{SPM}}$ with the minimum MSL achievable by measurements, $\mathcal{L}_{\mathrm{min}}$.

We introduce a figure of merit for measurement incompatibility of a given Bayesian multiparameter estimation problem as follows:
\begin{equation}
    \mathcal{I} \coloneq 
    \frac{ \mathcal{L}_{\mathrm{min}}  }{ \mathcal{L}_{\mathrm{SPM}} } - 1\geq 0 \, ,
    \label{eq:incompatibility-def}
\end{equation}
where the inequality follows from the SPM lower bound in Eq.~\eqref{eq:SPM_lower_bound}.
A similar figure of merit has been studied also in the framework of local QET~\cite{Belliardo2021}.
In the following sections, we shall derive upper and lower bounds on $\mathcal{I}$ by obtaining corresponding bounds on the unknown $\mathcal{L}_{\mathrm{min}}$.

\subsection{Upper bounds}

The incompatibility quantifier $\mathcal{I}$ can be upper bounded by considering specific POVMs and estimators, since the corresponding MSL is, by definition, never smaller than the achievable minimum. 

\subsubsection{Pretty good measurement}

Taking inspiration from hypothesis testing, we introduce the square root measurement, also known as the pretty good measurement (PGM)~\cite{Belavkin1975a,Belavkin1975,Kholevo1979,Hausladen1994}:
\begin{equation}
  M_{\mathrm{PGM}} (\boldsymbol{\tilde{\theta}}) \coloneq p(\boldsymbol{\tilde{\theta}})\, \rho_0^{-\frac{1}{2}}\,  \rho(\boldsymbol{\tilde{\theta}})\, \rho_0^{-\frac{1}{2}}.
  \label{eq:PGMdef}
\end{equation}
This measurement is usually introduced for discriminating among discrete hypotheses, where the Barnum--Knill theorem guarantees the ``pretty good'' performance~\cite{Barnum2002}.
Recently, this theorem has been extended to continuous hypotheses~\cite{Mishra2025} and applied to prove that the MSL of the PGM is within a factor two of the optimal $\mathcal{L}_{\mathrm{min}}$.
While this remarkable result establishes the usefulness of the PGM for parameter estimation, it does not shed light on the issue of measurement incompatibility, i.e., understanding the discrepancy between $\mathcal{L}_{\mathrm{min}}$ and $\mathcal{L}_{\mathrm{SPM}}$.
This will be the precise topic of the next subsection. 
In the remainder of this subsection, we set the stage by deriving the expression for the MSL achieved by the PGM.

Before proceeding, it is important to emphasise that the PGM is a POVM with outcomes in the parameter space and therefore already incorporates a classical estimator, which may not be optimal.
Nonetheless, it yields simple expressions for the MSL that are easy to interpret and compare with the baseline provided by the SPM bound.
In Sec.~\ref{subsubsec:PGMstar} we will show how the PM estimator---the optimal choice---can achieve a better MSL, at the expense of having a more involved expression.

The MSL of the PGM can be computed from Eq.~\eqref{eq:msl_compact}, by first evaluating the operators in Eqs.~\eqref{eq:measurement-moments} as
\begin{subequations}
\begin{align}
    M_{\text{PGM},i} &= \rho_0^{-\frac{1}{2}} \bar{\rho}_i \rho_0^{-\frac{1}{2}}, \\ 
    M_{\text{PGM},{ij}} &= \rho_0^{-\frac{1}{2}} \left[ \int  \! \mathrm{d} \tilde \params \, p(\tilde \params)\, \rho(\tilde\params)\, \tilde \theta_i \tilde \theta_j \right]  \rho_0^{-\frac{1}{2}}  .
\end{align}
\end{subequations}
Noticing that, from cyclicity of the trace,
\begin{equation}
L^{ij} \Tr [ \rho_0\,M_{\text{PGM},{ji}} ] = \int \mathrm{d}\params\, p(\params)\, \boldsymbol{\theta}^T L\, \boldsymbol{\theta} = \lambda , 
\end{equation}
since $\rho_0$ is cancelled by the two factors $\rho_0^{-\frac{1}{2}}$ and $\Tr[\rho(\tilde{\params})] = 1$, we obtain
\begin{equation}
\begin{aligned}
    \mathcal{L}_{\mathrm{PGM}} 
    & = 2 \left( \lambda - L^{ij} \Tr \left[ \bar{\rho}_j \rho_0^{-\frac{1}{2}} \bar{\rho}_i \rho_0^{-\frac{1}{2}} \right] \right) \\
    & = 2 \left( \lambda - L^{ij} \Tr \left[ \bar{\rho}_j E_{\rho_0,\mathrm{sq}}^{-1}[\bar{\rho}_i]  \right] \right) \, 
    \label{eq:MSL_PGM_inner} .
\end{aligned}
\end{equation}
In the second line, we have expressed the MSL in terms of the inverse of the map 
\begin{equation}
E_{\rho,\mathrm{sq}}[X] \coloneq \sqrt{\rho} X \sqrt{\rho} .
\label{eq:E-map-sqrt}
\end{equation}
This map, like \(E_{\rho,S}\) in Eq.~\eqref{eq:E-map-SLD}, is one of the infinitely many inequivalent ways of introducing a superoperator representing ``multiplication by $\rho$''.
This plurality is due to noncommutativity in quantum theory, and indeed for diagonal $\rho$ and $X$, i.e., embedding classical probability into quantum objects, they all correspond to an element-wise multiplication of probability vectors.
Such maps are fundamental mathematical objects in the theory of monotone metrics~\cite{Petz1996a,Hayashi2017c,Scandi2025}.

\subsubsection{Fundamental incompatibility upper bound}
\label{subsubsec:2LSPMupperbound}

A crucial result for us in the theory of monotone metrics is that the inner product obtained from the inverse of  \( E_\rho^S \) is the minimal one, implying the inequality
\begin{equation}
\label{eq:sym_vs_sqrt_monotone_ineq}
    \Tr(  X E_{\rho,\mathrm{S}}^{-1} \left[ X \right] ) \leq \Tr(  X E_{\rho,\mathrm{sq}}^{-1} \left[ X \right] ) \, .
\end{equation}
Notice that the two sides of the inequality are norms induced by different inner products between operators.
Since the weight matrix is symmetric and positive semidefinite, it admits a (non-unique) decomposition as a sum of rank-one outer products: $L = \sum_k \boldsymbol{v}_k \boldsymbol{v}_k^T$ for some real vectors $\boldsymbol{v}_k$.
This, together with linearity of $E_{\rho_0,\mathrm{sq}}^{-1}$, allows to rewrite the second term of Eq.~\eqref{eq:MSL_PGM_inner} as the sum of norms of the Hermitian operators $v_k^i \bar{\rho}_i$: 
\begin{equation}
\begin{split}
L^{ij} \Tr\left[ \bar{\rho}_j E_{\rho_0,\mathrm{sq}}^{-1}[\bar{\rho}_i]  \right] = 
\sum_k & \Tr \left[ ( v_k^i \bar{\rho}_i ) E_{\rho_0,\mathrm{sq}}^{-1}[ ( v_k^j \bar{\rho}_j ) ]  \right] .
\end{split}
\end{equation}
Plugging this identity into the second term of Eq.~\eqref{eq:MSL_PGM_inner} and applying the inequality~\eqref{eq:sym_vs_sqrt_monotone_ineq} to each summand, we prove the first inequality in the following chain
\begin{equation}
    2 \mathcal{L}_{\mathrm{SPM}} \geq \mathcal{L}_{\mathrm{PGM}} \geq \mathcal{L}_{\mathrm{min}} \geq \mathcal{L}_{\mathrm{SPM}} ,
    \label{eq:2LSPM-upper-bound}
\end{equation}
where the second inequality holds by definition and the third is the SPM lower bound in Eq.~\eqref{eq:SPM_lower_bound}, which neglects potential incompatibility effects.
Thus, we have succeeded in bounding the maximal increase in the MSL due to measurement incompatibility, which can equivalently be written as an upper bound on the incompatibility figure of merit:
\begin{equation}
    0 \leq \mathcal{I} \leq  1  .
    \label{eq:I-between-0-and-1}
\end{equation}

We remark that the upper bound in Eq.~\eqref{eq:2LSPM-upper-bound} is our second and one of our key results. 
Interestingly, this mirrors a well-known result in local QET: the HCRB is at most twice the Helstrom (SLD) CRB~\cite{Tsang2019,Carollo2019}.
Notice, however, that the HCRB is in general attainable only asymptotically with collective measurements on multiple copies.
Hence, only the fundamental asymptotic impact of measurement incompatibility is limited by the upper bound on the HCRB~\cite{Belliardo2021,Das2024}.
On the other hand, our result is obtained in a Bayesian estimation setting, and holds for a single shot.

\subsubsection{Pretty good measurement with optimal estimator}
\label{subsubsec:PGMstar}

While generally valid, the bounds derived above can be trivial, i.e., less informative than the \emph{a priori} MSL (cf. App.~\ref{app:prior-loss}):
$\mathcal{L}_{\mathrm{prior}} \leq \mathcal{L}_{\mathrm{PGM}}$, where
\begin{equation}
\mathcal{L}_{\mathrm{prior}} = \lambda - \boldsymbol{\mu}^T\,L\,\boldsymbol{\mu}, 
\quad \mu_i = \int \! \mathrm{d} \params\, p(\params)\, \theta_i.
\label{eq:prior-loss}
\end{equation}
Such a behaviour may appear counter-intuitive, since measuring a system is supposed to increase the information about the parameters' values.
The intuition is re-established if one realises that this happens only because of a suboptimal choice of the classical estimator in the construction of the PGM bound.
Similar situations may appear also in quantum state discrimination, i.e., for discrete hypotheses.
Indeed, it has been shown that, for specific ensembles, guessing the hypothesis that maximises the prior probability is optimal and the probability of success cannot be improved upon by performing measurements on the system~\cite{Hunter2003,Barnett2009a}.

Instead of viewing the PGM as a POVM already equipped with a built-in estimator---one not specifically tailored to our quadratic loss function---we now drop the interpretation of its outcomes as estimates.
For clarity, the POVM elements in Eq.~\eqref{eq:PGMdef} can be rewritten with a distinct label for the measurement outcomes as
\begin{equation}
    M_{\mathrm{PGM}}(\boldsymbol{x}) = p(\boldsymbol{x})\, \rho_0^{-\frac{1}{2}}\, \rho(\boldsymbol{x})\, \rho_0^{-\frac{1}{2}},
\end{equation}
where each $\boldsymbol{x}$ is still a vector in the hypothesis space $\Omega$, though it is not directly used as an estimator.
Instead, the measurement outcomes are processed through the optimal PM estimator in Eq.~\eqref{eq:classical_PM_estimator}, which is constructed from the likelihood function $p(\boldsymbol{x} | \params) = \Tr[ \rho(\params)\, M_{\mathrm{PGM}}(\boldsymbol{x})]$.
From a straightforward calculation, we obtain the MSL
\begin{equation}
\label{eq:MSL_PGMstar}
\begin{split}
\mathcal{L}_{\mathrm{PGM}^*} \coloneq \lambda 
- L^{ij} \int \mathrm{d}\boldsymbol{x} \, &p(\boldsymbol{x}) \,
    \Tr\left[ 
        \rho(\boldsymbol{x}) \rho_0^{-\frac{1}{2}} \bar{\rho}_j \rho_0^{-\frac{1}{2}} 
    \right] \\
    &\cdot
    \Tr\left[ 
        \rho(\boldsymbol{x}) \rho_0^{-\frac{1}{2}} \bar{\rho}_i \rho_0^{-\frac{1}{2}}
    \right].
\end{split}
\end{equation}
Contrary to the standard PGM, this quantity is guaranteed to be no greater than $\mathcal{L}_{\mathrm{prior}}$, which can be seen explicitly by applying Jensen's inequality to the second term of Eq.~\eqref{eq:MSL_PGMstar}, since the function of $\rho(\params)$ is quadratic and convex.
Whilst for an arbitrary POVM with the same estimator---which is guaranteed to be smaller than \( \mathcal{L}_{\mathrm{prior}} \)---the MSL may be larger than $\mathcal{L}_{\mathrm{PGM}}$.
Note that, as discussed, for our choice of quadratic loss we have a formal guarantee that the PM is the optimal classical estimator, but there is numerical evidence that it also leads to a better fidelity than the standard PGM approach in state estimation~\cite{Quadeer2024}.

Tightening the previous inequalities, by taking into account that they may be trivial, and adding this new tighter upper bound, we obtain
\begin{equation}
\begin{aligned}
    \min\!\left[ 2\mathcal{L}_{\mathrm{SPM}},\, \mathcal{L}_{\mathrm{prior}} \right]
    &\geq \min\!\left[ \mathcal{L}_{\mathrm{PGM}},\, \mathcal{L}_{\mathrm{prior}} \right] \\
    &\geq \mathcal{L}_{\mathrm{PGM}^*} \\
    &\geq \mathcal{L}_{\mathrm{min}} \\
    &\geq \mathcal{L}_{\mathrm{SPM}} \, .
    \label{eq:upper-bounds-result}
\end{aligned}
\end{equation}
The corresponding bounds on $\mathcal{I}$ are
\begin{equation}
\begin{aligned}
    0 &\leq \mathcal{I} \\
    &\leq \mathcal{I}_{\mathrm{PGM}^*} \\
    &\leq \min\!\left[ \mathcal{I}_{\mathrm{PGM}},\, \mathcal{I}_{\mathrm{prior}} \right] \\
    &\leq \min\!\left[ 1,\, \mathcal{I}_{\mathrm{prior}} \right],
    \label{eq:upper-bounds-incomp}
\end{aligned}
\end{equation}
where $\mathcal{I}_{\mathrm{PGM}}$, $\mathcal{I}_{\mathrm{PGM^*}}$, and $\mathcal{I}_{\mathrm{prior}}$ are defined as in Eq.~\eqref{eq:incompatibility-def}.
The first three inequalities in Eq.~\eqref{eq:upper-bounds-result} (the last three inequalities in Eq.~\eqref{eq:upper-bounds-incomp}) constitute our third result.

\subsection{Lower bounds}\label{sec:lower-bounds}

Having established an upper bound on the MSL to characterise the worst-case impact of measurement incompatibility, as well as a tighter version, our aim is now to tighten the lower bound in Eq.~\eqref{eq:I-between-0-and-1} for a given model. To this end, we seek lower bounds on the MSL that are tighter than $\mathcal{L}_{\mathrm{SPM}}$ in the presence of non-commuting SPM operators. In this section, we focus primarily on the tightest known lower bound on the MSL available in the literature, the Nagaoka–Hayashi (NH) bound, \(\mathcal{L}_{\mathrm{NH}}\), recently introduced in Bayesian QET by Suzuki~\cite{Suzuki2024a}.

\subsubsection{Nagaoka--Hayashi bound}

To derive this bound, one starts by noticing that any choice of POVM and estimators satisfies the following inequality~\cite[Lemma 1]{Suzuki2024a},
\begin{equation}
 \sum_{ij} |i \rangle \langle j| \otimes M_{ij} \succeq \sum_{ij} |i \rangle \langle j| \otimes M_i M_j \, ,
 \label{eq:lemma1-Suzuki}
\end{equation}
in terms of the first and second moment operators introduced in  Eq.~\eqref{eq:measurement-moments}.
Note that an additional vector space of dimension $d$ (number of parameters) has been introduced, but this is not a physical Hilbert space.
While we use the bra--ket notation, $\ket{k}$ simply stands for the standard column basis vector $e_k$ of $\mathbbm{R}^{d}$ (we have also temporarily reintroduced an explicit summation for extra clarity).
The two sides of Eq.~\eqref{eq:lemma1-Suzuki} are block matrices: the left-hand side can be interpreted as an operator-valued second moment matrix, while the right-hand side as an outer product of operator-valued first moment vectors.

Since the MSL in Eq.~\eqref{eq:msl_compact} is expressed only in terms of these operators, a lower bound is obtained by treating the Hermitian operators $M_{ij}=M_{ji}$ and $M_i$ as \emph{independent} variables satisfying Eq.~\eqref{eq:lemma1-Suzuki}: 
\begin{align}
\label{eq:NHbound}
    \mathcal{L}_{\mathrm{NH}} \coloneq \lambda  +  & \min_{\{M_{ij} , M_i\}}  L^{ij}\,\operatorname{Tr}[\rho_0 M_{ji} - 2 \bar{\rho}_j M_i] \\
    & \, \, \mathrm{subject~to~}  \mathrm{Eq.~}\eqref{eq:lemma1-Suzuki} \, . \nonumber 
\end{align}
This means that, for a given set of \( M_{ij} \) and \( M_i \) attaining the minimum, there is no guarantee that an underlying POVM $M(x)$ exists from which they can be derived, according to Eqs.~\eqref{eq:measurement-moments}.
Consequently, the bound may be loose.
However, despite not \emph{fully} capturing measurement incompatibility, the NH bound can be shown to be tighter than the SPM one.
This follows by noticing that Eq.~\eqref{eq:lemma1-Suzuki} implies 
\begin{equation}
    L^{ij} \Tr[ \rho_0 M_{ji} ] \geq L^{ij} \Tr[ \rho_0 M_j M_i ],
\end{equation}
which is a weaker constraint.
Applying this inequality to Eq.~\eqref{eq:NHbound} one obtains a relaxation of the original problem that can be solved analytically to obtain $\mathcal{L}_{\mathrm{SPM}}$.

The bound $\mathcal{L}_{\mathrm{NH}}$ formally mirrors the NH Cramér-Rao bound (NHCRB) in local QET~\cite{Conlon2020}, and analogously it can be evaluated by solving a semidefinite program (SDP)~\cite{Suzuki2024a}.
Notice, however, that the upper bound in Eq.~\eqref{eq:2LSPM-upper-bound} that holds in the Bayesian case has no direct counterpart in the local case, since the NHCRB can be much larger than twice the Helstrom (SLD) CRB, and this gap can grow with the dimension of the Hilbert space and number of parameters~\cite{Das2024}.

For two parameters, the optimisation over the second moment operators $M_{11}$, $M_{12}$ and $M_{22}$ can be solved exactly.
For the sake of clarity, but without loss of generality\footnote{
Similarly to local QET, the scalar NH bound can be recast to an equivalent problem with $L=\id$ and suitably redefined first moment operators $\bar{\rho}_{i}' = \sum_j (\sqrt{L})^{ij} \bar{\rho}_j$.
The explicit expression for a general 
weight matrix is also reported in Ref.~\cite[Eq.~(20)]{Suzuki2024a}.}, we present this expression for an identity weight matrix:
\begin{equation}
\begin{split}
    \mathcal{L}_{\mathrm{NH}} \overset{L=\id_2}{=} \lambda + \min_{M_1,M_2} \{ \Tr\left[ \rho_0 (M_1^2 + M_2^2) \right]  \\ 
    + \left\Vert \sqrt{\rho_0} [ M_1,M_2] \sqrt{\rho_0} \right\Vert_1  - 2 \Tr \left[ \bar{\rho}_1 M_1 + \bar{\rho}_2 M_2 \right] \} \, ,
\end{split}
\label{eq:Nagaoka2par}
\end{equation}
where \( \Vert A \Vert_1  = \Tr[ \sqrt{ A^\dag A } ] \) denotes the trace norm, i.e., the sum of the singular values.
This expression can be derived from Eq.~\eqref{eq:NHbound}, following, e.g., Ref.~\cite[Supplementary Note 2]{Conlon2020}.
This form is the one originally obtained by Nagaoka for local two-parameter estimation~\cite{Nagaoka1989}.

Further considering a single qubit and full-rank $\rho_0$ (the only nontrivial qubit case), the NH bound in Eq.~\eqref{eq:Nagaoka2par} is in fact attainable, i.e.,
$\mathcal L_{\mathrm{NH}}=\mathcal L_{\mathrm{min}}$, because Nagaoka's argument carries over to the Bayesian setting; we show this
in App.~\ref{app:NH_tight_qubit_d2}.
The proof is constructive and yields an explicit optimal POVM, reported in App.~\ref{app:NH_tight_qubit_d2:povm}.
Moreover, this case can be solved in closed form: the remaining optimisation over $M_1,M_2$ in Eq.~\eqref{eq:Nagaoka2par} is evaluated in App.~\ref{app:NH_qubit_closed_form}.

Overall, we have the following inequalities:
\begin{equation}
\mathcal{L}_{\mathrm{min}} \geq \mathcal{L}_{\mathrm{NH}} \geq \mathcal{L}_{\mathrm{SPM}} \, .
\end{equation}
In turn, this restricts the allowed range of the incompatibility figure of merit for a given model:
\begin{equation}
    \mathcal{I}_{\mathrm{NH}} \leq \mathcal{I} \leq  1 \, ,
\end{equation}
where $\mathcal{I}_{\mathrm{NH}}$ is defined analogously to Eq.~\eqref{eq:incompatibility-def}.
Moreover, by combining this NH bound on incompatibility with the tightest upper bound in Eq.~\eqref{eq:upper-bounds-result}, we obtain a tight range for the possible values of incompatibility in Bayesian multiparameter quantum estimation:
\begin{equation}
    \mathcal{I}_{\mathrm{NH}} \leq \mathcal{I} \leq \mathcal{I}_{\mathrm{PGM}^*}.
    \label{eq:incomp-range}
\end{equation}
This constitutes our fourth result.

\subsubsection{Other lower bounds}

We conclude this section by highlighting other lower bounds on the MSL which, although less tight---and mostly derived using techniques from local QET---can be easier to compute and may therefore remain useful in practice.

We first mention the Holevo bound, here denoted as $\mathcal{L}_{\mathrm{H}}$, which is a Bayesian version of the HCRB~\cite{Suzuki2024a,Zhang2024g}.
It can be obtained by loosening the constraints in Eq.~\eqref{eq:lemma1-Suzuki} taking the expectation value with respect to $\rho_0$, thereby obtaining a minimisation problem with an inequality constraint involving $d{\times }d$ matrices.
The HCRB sits between the SPM and NH bounds: $\mathcal{L}_{\mathrm{NH}} \geq \mathcal{L}_{\mathrm{H}} \geq \mathcal{L}_{\mathrm{SPM}}$.
While in local QET the HCRB has a fundamental meaning, being asymptotically attainable, the Bayesian version does not seem to share the same significance\footnote{Note that such a Bayesian Holevo-type bound is different from the so-called Bayesian CRBs obtained by averaging the HCRB over the prior~\cite{Gill2008}, conjectured to be attainable in the large sample limit of Bayesian estimation~\cite[Sec.~5.5]{demkowicz2020multi}; see the discussion in Sec.~\ref{sec:discussion}.}.
The Bayesian Holevo bound may be computationally more efficient because, while it can be computed by solving an SDP as the NH bound, it involves positive semidefinite constraints in terms of matrices of smaller dimension~\cite{Suzuki2024a}.

Pursuing further the goal of identifying easily computable bounds, one may instead seek lower bounds that capture measurement incompatibility to some extent and can always be expressed in closed form, rather than indirectly as the outcome of an optimisation.
A whole class of such bounds exists, based on the idea of non-commutative PM operators, of which SPMs are just one example.
This is in full analogy with logarithmic derivative operators in local QET~\cite[Ch.~6]{Hayashi2017c}. 
In App.~\ref{app:matrix-lower-bounds}, we derive this family of lower bounds in a unified manner, employing the theory of monotone metrics, and generalising the approach of Ref.~\cite{rubio2020bayesian}.
Here, we only mention the right posterior mean (RPM) bound 
\begin{subequations}
\begin{align}
    \mathcal{L}_{\mathrm{RPM}} &\coloneq \lambda - L^{ij} \mathcal{K}_{\mathrm{R},ji} +  \Vert \sqrt{L} \Im[\mathcal{K}_{\mathrm{R}}] \sqrt{L} \Vert_1 ,         \label{eq:RPMbound} \\
    \mathcal{K}_{\mathrm{R},ij} & \coloneq \Tr [ \bar{\rho}_i \rho_0^{-1} \bar{\rho}_j  ],
    \end{align}
\end{subequations}
written in terms of the complex matrix $\mathcal{K}_{\mathrm{R}}$.
Such a matrix is analogous to the right logarithmic derivative quantum Fisher information matrix in local QET.
This bound is less informative than the NH or Holevo ones, since $\mathcal{L}_{\mathrm{H}} \geq \mathcal{L}_{\mathrm{RPM}}$~\cite[Thm.~8]{Zhang2024g}. 
However, it can be tighter or looser than \( \mathcal{L}_{\mathrm{SPM}} \) depending on the specific model.
For example, it is always less tight than the SPM bound for $d=1$, where the latter is attainable.
Conversely, Holevo has shown the RPM bound to be attainable for a multiparameter problem with non-commuting SPM operators: estimation of the average position and momentum of a Gaussian state with a Gaussian prior~\cite{kholevo1978investigations}.

As a means of summary, we combine the upper bounds in Eq.~\eqref{eq:upper-bounds-result} with all the lower bounds discussed in this section to obtain the following hierarchy of bounds on the minimum MSL:
\begin{equation}
\begin{aligned}
    \min\!\left[ 2\mathcal{L}_{\mathrm{SPM}},\, \mathcal{L}_{\mathrm{prior}} \right]
    &\geq \min\!\left[ \mathcal{L}_{\mathrm{PGM}},\, \mathcal{L}_{\mathrm{prior}} \right] \\
    &\geq \mathcal{L}_{\mathrm{PGM}^*} \\
    &\geq \mathcal{L}_{\mathrm{min}} \\
    &\geq \mathcal{L}_{\mathrm{NH}} \\
    &\geq \mathcal{L}_{\mathrm{H}} \\
    &\geq \max\!\left[ \mathcal{L}_{\mathrm{SPM}},\, \mathcal{L}_{\mathrm{RPM}} \right].
    \label{eq:all-bounds-msl}
\end{aligned}
\end{equation}

Correspondingly, the hierarchy of bounds on measurement incompatibility is
\begin{equation}
\begin{aligned}
    \max\!\left[ 0,\, \mathcal{I}_{\mathrm{RPM}} \right]
    &\leq \mathcal{I}_{\mathrm{H}} \\
    &\leq \mathcal{I}_{\mathrm{NH}} \\
    &\leq \mathcal{I} \\
    &\leq \mathcal{I}_{\mathrm{PGM}^*} \\
    &\leq \min\!\left[ \mathcal{I}_{\mathrm{PGM}},\, \mathcal{I}_{\mathrm{prior}} \right] \\
    &\leq \min\!\left[ 1,\, \mathcal{I}_{\mathrm{prior}} \right],
    \label{eq:all-bounds-incomp}
\end{aligned}
\end{equation}
where $\mathcal{I}_{\mathrm{RPM}}$ and $\mathcal{I}_{\mathrm{H}}$ are defined as in Eq.~\eqref{eq:incompatibility-def}.

\subsection{Approaching the minimum MSL numerically}
\label{sec:seesaw-numerics}

While the bounds derived above enable efficient estimation of the incompatibility range, the exact value of the minimum MSL $\mathcal{L}_{\min}$ typically requires numerical optimisation over POVMs.
A practical approach, introduced in Ref.~\cite{Bavaresco2024designing} in the more general setting of channel estimation, is an iterative \emph{see-saw} (alternating optimisation) method.
Here, we briefly summarise the variant relevant to our setting of parameter-dependent families of states.

We restrict, without loss of generality, to POVMs with a finite number of outcomes, denoted as $\{M_n\}_{n=1}^{N}$; see the definition in Eq.~\eqref{eq:discrete-POVM}, with associated estimator values $\{\boldsymbol{\tilde{\theta}}_n\}_{n=1}^{N}$.
For the quadratic loss in Eq.~\eqref{eq:square_loss}---or its symmetry-remapped version in Eq.~\eqref{eq:symmetry-msl}---the MSL can be written as a constant term depending only on the prior moments plus a term that is linear in the POVM elements for fixed estimators.
Concretely, defining $\tilde{\theta}_{n,i}$ as the $i$-th component of the estimator associated with outcome $n$, one can rewrite
\begin{equation}
\mathcal{L} = \lambda - \sum_{n=1}^{N} \Tr\!\big[M_n\,X_n(\boldsymbol{\tilde{\theta}}_n)\big],
\end{equation}
where each $X_n(\boldsymbol{\tilde{\theta}}_n)$ is a Hermitian operator that depends only on the zeroth and first state moments $\rho_0$ and $\{\bar\rho_i\}$ in Eqs.~\eqref{eq:state-moments} and \eqref{eq:first-state-moment}, and on the weight matrix $L$.
The explicit expression is obtained by inserting the definition of the measurement moments for a discrete-outcome POVM
$M_i=\sum_n \tilde{\theta}_{n,i} M_n$ and $M_{ij}=\sum_n \tilde{\theta}_{n,i}\tilde{\theta}_{n,j} M_n$ into Eq.~\eqref{eq:msl_compact}, which yields
\begin{equation}
X_n(\tilde{\boldsymbol\theta}_n)
= 2\,L^{ij}\,\tilde{\theta}_{n,i}\,\bar\rho_j
- L^{ij}\,\tilde{\theta}_{n,i}\tilde{\theta}_{n,j}\,\rho_0.
\label{eq:seesaw-Xa}
\end{equation}
Therefore, for fixed $\{\boldsymbol{\tilde{\theta}}_n^{(t)}\}$, where $t$ labels each iteration, the optimisation over the measurement reduces to a SDP:
\begin{equation}
\max_{\{M_n\}}\;\sum_{n=1}^{N} \Tr\!\big[M_n\,X_n(\tilde{\boldsymbol\theta}_n^{(t)})\big]
\quad\text{s.t.}\quad
M_n\succeq 0,\ \sum_{n=1}^{N} M_n = \id.
\label{eq:seesaw-sdp}
\end{equation}
This step is convex and can be solved to global optimality for the chosen number of outcomes $N$.
For a $D$-dimensional system, at most $N=D^2$ outcomes are needed~\cite{Bavaresco2024designing}.

For fixed POVM $\{M_n^{(t)}\}$, obtained by solving the SDP above, the optimal classical post-processing for quadratic loss is the PM estimator in Eq.~\eqref{eq:PMestimator} and no optimisation is needed; for different loss functions, a second optimisation is generally needed~\cite{Bavaresco2024designing}.
In our notation, this update can be expressed directly in terms of the state moments:
\begin{equation}
\tilde{\theta}_{n,i}^{(t+1)} = \frac{\Tr\!\big[M_n^{(t)}\,\bar\rho_i\big]}{\Tr\!\big[M_n^{(t)}\,\rho_0\big]},
\label{eq:seesaw-pm-update}
\end{equation}
and analogously after the symmetry remapping $\boldsymbol\theta\mapsto\boldsymbol f(\boldsymbol\theta)$.

Iterating Eqs.~\eqref{eq:seesaw-sdp} and \eqref{eq:seesaw-pm-update} yields a sequence of feasible strategies whose MSL is non-increasing at every step.
The procedure is not guaranteed to find the global optimum in general---the joint optimisation over POVMs and estimators is non-convex---but in practice it provides a constructive upper bound on $\mathcal{L}_{\min}$.
To mitigate the risk of converging to suboptimal stationary points, one may perform multiple random initialisations of the estimators and keep the best solution.
Finally, candidate solutions produced by the see-saw could also be checked against the optimality conditions in Eq.~\eqref{eq:msl-opt-condition}, providing an \emph{a posteriori} consistency test.

This gives (as any POVM would) an upper bound on incompatibility (\(\mathcal{I}_{\mathrm{SSW}}\)).
However, iteratively solving the SDPs is an additional layer of computational complexity, potentially requiring a scaling---in the number of POVM elements to solve for---with the square of the state-space dimension.

\section{Applications}\label{sec:applications}

In the following sections we evaluate the measurement incompatibility range in Eq.~\eqref{eq:incomp-range}, as well as some of the less tight bounds derived, for a range of examples.

\subsection{Discrete quantum phase imaging}\label{sec:imaging}

Consider a $(d{+}1)$-mode interferometer, where $d$ parameters $\boldsymbol{\theta}$ for a set of unknown phases $\boldsymbol{\Theta}$ are encoded via the unitary operator $\mathrm{e}^{-i N^j \theta_j}$. 
Here, $N^j$ is the number operator of the $j$th mode, and the zeroth mode serves as a reference beam~\cite{jarzyna2012quantum}.
This configuration models discrete quantum imaging and has been extensively explored in both local and Bayesian frameworks~\cite{macchiavello2003optimal,humphreys2013quantum,Albarelli2019c,rubio2020bayesian,gebhart2021bayesian}.
In this context, a common probe state is the \( n \)-photon generalised N00N state 
\begin{equation}
    \ket{\psi(\theta)} = \frac{1}{\sqrt{d + \alpha^2}}
    \left(\alpha e^{-i n\theta_0}\ket{n_0} + \sum_{i=1}^d e^{-i n\theta_i}\ket{n_i}\right),
    \label{eq:imaging:state}
\end{equation}
with $\alpha \in \mathbb{R}$ and
\begin{equation}
    \ket{n_i} := \ket{0}_0 \otimes \cdots \otimes \ket{0}_{i-1} \otimes \ket{n}_i \otimes \ket{0}_{i+1} \otimes \cdots \otimes \ket{0}_d
\end{equation}
representing a state with $n$ photons in the $i$-th mode, such that \( \sum\limits_j N^j \ket{\psi(\theta)} = n \ket{\psi(\theta)} \).

To address this problem using the MSL, we restrict the hypothesis space $\Omega$ so that the phases are sufficiently localised to behave as location parameters, but remaining within a global regime.
Specifically, we employ a flat prior $p(\boldsymbol{\theta}) = \left( \frac{n}{2\pi} \right)^d$, with $\theta_i \in [-\pi/n, \pi/n]$ and $n \geq 4$, so that the largest prior width for the $i$-th parameter is $\pi/2$.
For this configuration---and choosing a normalised $L$ with equal weights, i.e., $L = \mathbb{I}/d$---the average state, SPM operators, and SPM bound are, respectively~\cite{rubio2019quantum},
\begin{gather}
    \rho_0 = \frac{1}{d + \alpha^2} \left( \alpha^2 \ketbra{n_0}{n_0} + \sum_{i=1}^d\ketbra{n_i}{n_i} \right), \label{eq:application:imaging:rho0}
\\
    \bar{\rho}_j = \frac{-i\alpha}{n(d + \alpha^2)} \left(\ketbra{n_j}{n_0} - \ketbra{n_0}{n_j}\right),
\\
    S_j = \frac{-2i\alpha}{n(1 + \alpha^2)}
\left(\ketbra{n_j}{n_0} - \ketbra{n_0}{n_j}\right),
\end{gather}
and 
\begin{equation}
    \mathcal{L}_{\mathrm{SPM}} = \frac{1}{n^{2}}
\left[
\frac{\pi^{2}}{3}
- \frac{4\alpha^{2}}{(1+\alpha^{2})(d+\alpha^{2})}
\right].
\label{eq:spm-bound-imaging}
\end{equation}

Two observations are in order. 
First, the Heisenberg scaling in Eq.~\eqref{eq:spm-bound-imaging} is achievable in a global setting only because the chosen prior width scales as $\mathord{\sim} 1/n$.
Indeed, since all first moments of the prior vanish for this problem, the \emph{a priori} MSL is given by (cf.~Eq.~\eqref{eq:prior-loss})
\begin{equation}
    \mathcal{L}_{\mathrm{prior}} = \lambda = \frac{\pi^2}{3 n^2},
\end{equation}
which already exhibits Heisenberg scaling; this is consistent with the observations in Refs.~\cite{hall2012does, giovannetti2012subheisenberg, tsang2012zivzakai, pezze2013subheisenberg, rubio2018nonasymptotic,branford2021average}.

Secondly, and more importantly for our purposes, the SPM operators do not commute:
\begin{equation}
    [S_j, S_k] = \frac{4 \alpha^2}{n^2 (1 + \alpha^2)^2}(\ketbra{n_j}{n_k} - \ketbra{n_k}{n_j}).
\end{equation}
As such, the SPM bound is not tight, and we can therefore expect incompatibility to arise in this system.

To quantify the amount of incompatibility, we first need to calculate the bounds on the minimum MSL.  
We start with the PGM bound, given in Eq.~\eqref{eq:MSL_PGM_inner}.
Since
\begin{equation}
    \bar{\rho}_j \, \rho_0^{-\frac{1}{2}} = \frac{-i}{n\sqrt{d+\alpha^2}}
    \left(\ketbra{n_j}{n_0} - \alpha \ketbra{n_0}{n_j} \right),
\end{equation}
and so 
\begin{equation}
    \Tr [ \bar{\rho}_j \, \rho_0^{-\frac{1}{2}} \, \bar{\rho}_i \, \rho_0^{-\frac{1}{2}} ] 
    = \frac{2\alpha}{n^2 (d + \alpha^2)} \delta_{ji},
\end{equation}
Eq.~\eqref{eq:MSL_PGM_inner} yields
\begin{equation}
    \mathcal{L}_{\mathrm{PGM}} = \frac{2}{n^2} \left( \frac{\pi^2}{3} - \frac{2\alpha}{d + \alpha^2} \right).
\end{equation}
However, this upper bound is trivial, i.e., $\mathcal{L}_{\mathrm{PGM}} > \mathcal{L}_{\mathrm{prior}}$, and consequently so is $2\mathcal{L}_{\mathrm{SPM}}$.
This follows from showing that $\mathcal{L}_{\mathrm{PGM}} - \mathcal{L}_{\mathrm{prior}}$ is strictly positive by virtue of the inequality
\begin{equation}
    \frac{\pi^2}{3} > \max_{\alpha} \frac{4\alpha}{d + \alpha^2} = \frac{2}{\sqrt{d}},
\end{equation}
valid for all $d$. 

In view of this, we proceed to calculate the PGM bound enhanced by the PM estimator, given in Eq.~\eqref{eq:MSL_PGMstar}, which is guaranteed to be smaller than or equal to the \emph{a priori} MSL.
First, calculate
\begin{equation}
\rho_0^{-\frac{1}{2}} \, \bar{\rho}_j \, \rho_0^{-\frac{1}{2}} = \frac{i}{n} (\ketbra{n_0}{n_j}-\ketbra{n_j}{n_0}),
\end{equation}
so that
\begin{equation}
    \Tr [ \rho(\params) \, \rho_0^{-\frac{1}{2}} \, \bar{\rho}_j \, \rho_0^{-\frac{1}{2}} ] = \frac{2\alpha \sin(n\theta_j)}{n(d+\alpha^2)}.
\end{equation}
Since the following integrals hold:
\begin{equation}
    \int\limits_{-\pi/n}^{\pi/n}  \mathrm{d}\theta_j \, \sin(n\theta_j) 
        = 0, \!
    \int\limits_{-\pi/n}^{\pi/n}  \mathrm{d}\theta_j \, \sin^2(n\theta_j) 
        = \frac{\pi}{n},
\end{equation}
Eq.~\eqref{eq:MSL_PGMstar} leads to
\begin{equation}
    \mathcal{L}_{\mathrm{PGM}^*} = \frac{1}{n^2} \left( \frac{\pi^2}{3} -  \frac{2 \alpha^{2}}{\left(d + \alpha^{2}\right)^{2}} \right).
\end{equation}
This is our tightest upper bound on the minimum MSL. 

\begin{figure}[!tb]
    \includegraphics[trim={0cm 0cm 0cm 0cm},clip,width=\linewidth]{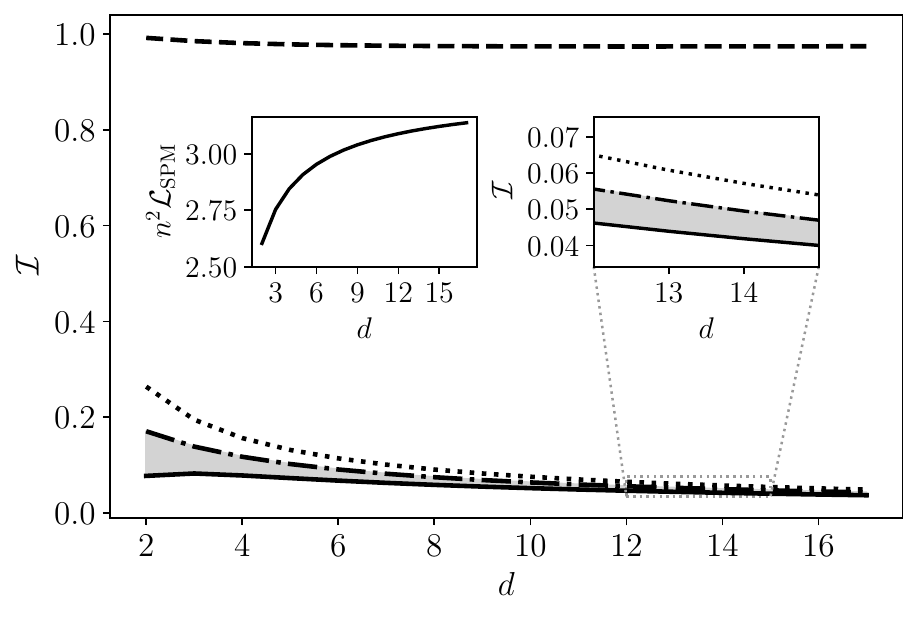}
    \caption{
    Measurement incompatibility bounds from
    the NH bound (solid), the PGM bound enhanced by the PM estimator (dash--dot), the standard PGM bound (dashed), and the \emph{a priori} MSL (dotted)
    for a discrete quantum phase imaging protocol.
    The shaded region represents the range of incompatibility, as defined in Eq.~\eqref{eq:incomp-range}, as a function of the number of parameters $d$, using a generalised N00N state with $\alpha = d^{1/4}$ as the initial state.
    The first inset displays the reference MSL, $\mathcal{L}_{\mathsf{SPM}}$.
    %As observed,
    The incompatibility associated with the \emph{a priori} MSL decreases with the number of parameters, thereby fundamentally limiting the extent to which measurement incompatibility can influence the overall precision;
    a zoomed view for larger parameter values is shown in the second inset. 
    Furthermore, the incompatibility for the standard PGM bound is considerably larger than its \emph{a priori} MSL counterpart and thus trivial for this system.}
\label{fig:figure-1}
\end{figure}

We next turn to the lower bounds.
We begin with the simplest tractable bound beyond the SPM bound, namely the RPM bound in Eq.~\eqref{eq:RPMbound}. Using
\begin{equation}
    \bar{\rho}_j \, \rho_0^{-1} \, \bar{\rho}_k
    = \frac{1}{n^2 (d+\alpha^2)}\left(\ketbra{n_j}{n_k}
    + \alpha^2\delta_{jk}\ketbra{n_0}{n_0}\right),
\end{equation}
we obtain
\begin{equation}
    \mathcal{K}_{\mathrm{R},jk}
    = \frac{1+\alpha^2}{n^2 (d+\alpha^2)} \, \delta_{jk},
    \label{eq:K-imaging}
\end{equation}
and therefore
\begin{equation}
    \mathcal{L}_{\mathrm{RPM}}
    = \frac{1}{n^2}\left(\frac{\pi^2}{3}
    - \frac{1+\alpha^2}{d+\alpha^2}\right).
\end{equation}
However, it is straightforward to verify that $\mathcal{L}_{\mathrm{SPM}} \ge \mathcal{L}_{\mathrm{RPM}}$, consistent with the fact that $\mathcal{K}_{\mathrm{R}}$ in Eq.~\eqref{eq:K-imaging} is real-valued. Consequently, we shall instead evaluate the NH bound in Eq.~\eqref{eq:NHbound} using the numerical algorithm of Ref.~\cite{multipar2025code} to obtain a tighter lower bound on the minimum MSL.

We are now in a position to quantify measurement incompatibility. 
Building on the preceding results, Fig.~\ref{fig:figure-1} illustrates as a shaded region the range defined in Eq.~\eqref{eq:incomp-range}, showing its dependence on the number of parameters \(d\) and using the state with $\alpha = d^{1/4}$ that minimises Eq.~\eqref{eq:spm-bound-imaging}.
The upper boundary of this region is given by $\mathcal{I}_{\mathrm{PGM^*}}$ (dashed-dot line), while the lower boundary corresponds to $\mathcal{I}_{\mathrm{NH}}$ (solid line). 
For comparison, we also include the incompatibility quantifier associated with the \emph{a priori} MSL (dotted line) and the PGM bound (dashed line). 
As can be seen, the latter exceeds the former and is therefore trivial, as discussed above.
Finally, the reference value $\mathcal{L}_{\mathrm{SPM}}$ is shown in the first inset, likewise plotted as a function of $d$.

Fig.~\ref{fig:figure-1} confirms the presence of measurement incompatibility within this protocol, with values not exceeding $0.2$. 
This relatively low magnitude can initially be attributed to the proximity between $\mathcal{L}_{\mathrm{prior}}$ and $\mathcal{L}_{\mathrm{SPM}}$, evidenced by the dotted line. 
That is, within the Bayesian framework, the prior information imposes a finite and natural upper bound on the MSL, thereby limiting the maximum attainable relative precision gain.
This effect becomes especially evident as the number of parameters increases, leading to a progressive reduction in relative precision gain and leaving increasingly limited scope for incompatibility to exert a noticeable influence on the precision. 
Nevertheless, the second inset demonstrates that the protocol consistently exhibits a nonzero degree of measurement incompatibility.

Here we see a deviation from the results for this model in local QET~\cite{humphreys2013quantum,Pezze2017}, where the SLD bound is attainable---and no incompatibility is seen---due to the commuting generators which give rise to commuting SLD operators for pure state probes.
With pure states leading to a mixed-state quantum statistical model in the Bayesian framework,
the corresponding SPM operators do not directly commute and incompatibility is found as the SPM bound is unobtainable.
Nevertheless that incompatibility is small, and numerical evidence using the method described in Sec.~\ref{sec:seesaw-numerics} suggests it is at the lower end of the incompatibility window shown in Fig.~\ref{fig:figure-1} as the NH bound appears attainable with
\( \abs{\mathcal{I}_{\mathrm{NH}} - \mathcal{I}_{\mathrm{SSW}}} \leq 3{\times}10^{-4} \) for $2 \leq d \leq 10$~\cite{supplemental_code}.

Symmetries, including the commuting generators, of the physical configuration may yet be responsible for both the low incompatibility,
and apparent numeric saturability of the NH bound.
It is perhaps informative to note that, while lacking the operational meaning it has in local QET~\cite{ragy2016compatibility},
a Bayesian analogue of the so-called weak commutativity, 
\begin{equation}
    \Tr\left[\rho_0 [S_j, S_k]\right] = 0,
    \label{eq:bayes-wc}
\end{equation}
does still hold in this case, since
\begin{equation}
    \rho_0 [S_j, S_k]
    = \frac{4\alpha^2}{n^2 (1+\alpha^2)^2 (d+\alpha^2)}
      \bigl(\ketbra{n_j}{n_k} - \ketbra{n_k}{n_j}\bigr).
\end{equation}

\subsection{Phase and dephasing estimation}\label{sec:phase-dephasing}

While phase estimation serves as the archetypal problem in quantum estimation, the simultaneous estimation of phase and phase diffusion plays an analogous role in multiparameter quantum estimation. 
The simplest single-qubit incarnation of this model exhibits measurement incompatibility in local QET~\cite{vidrighin2014joint}, which however asymptotically disappears as many identical copies are measured collectively~\cite{roccia2018entangling}.
This corresponds to preparing identical and factorised probe states, whereas asymptotically large entangled probes generally allow to optimally estimate both parameters~\cite{kobus202asymptotically}.
Nevertheless, we consider here independent identical copies collectively measured, since this is a paradigmatic setting for investigating measurement incompatibility~\cite{vidrighin2014joint,roccia2018entangling,Yung2025}.

For an initial state $\rho_{\mathrm{in}}$, the single-qubit phase and phase-diffusion channels\footnote{The qubit can represent, e.g., a single-photon in optical interferometry, whereas the same setup for probe states with more photons is not equivalent to considering multiple copies of the single-photon state, but give rise to correlated dephasing~\cite{Genoni2011a,Szczykulska2017}.}, which commute, are defined as
\begin{subequations}
\begin{align}
\Lambda_1(\rho_{\mathrm{in}}) 
&= \mathrm{e}^{-i\frac{\theta_1}{2}\sigma_{z}} \,\rho_{\mathrm{in}}\, \mathrm{e}^{i\frac{\theta_1}{2}\sigma_{z}}, 
\label{eq:commuting_channels_a}\\
\Lambda_2(\rho_{\mathrm{in}}) 
&= \int\limits_{-\infty}^{\infty} \mathrm{d}x \, 
    \frac{\mathrm{e}^{-\frac{x^2}{4\theta_2^2}}}{2\sqrt{\pi}\,\theta_2}\,
    \mathrm{e}^{-i\frac{x}{2}\sigma_{z}} \,\rho_{\mathrm{in}}\, \mathrm{e}^{i\frac{x}{2}\sigma_{z}},
\label{eq:commuting_channels_b}
\end{align}
\end{subequations}
respectively; $\sigma_x$, $\sigma_y$ and $\sigma_z$ denote Pauli matrices.
The corresponding parameter-dependent state is 
\begin{equation}
    \rho(\params) = \frac{1}{2} (\mathbb{I}_2 + 
        \cos\theta_1 \mathrm{e}^{-\theta_2^2} \sigma_x -\sin\theta_1 \mathrm{e}^{-\theta_2^2} \sigma_y) \, .
        \label{eq:state-phase-diffusion}
\end{equation}

The MSL is not suitable for this problem.
Specifically, while $\theta_1 \in [-\frac{W_1}{2}, \frac{W_1}{2}]$ is a cyclic variable that can be treated as a location parameter when $W_1 \lesssim \pi/2$---see Sec.~\ref{subsec:estimation-performance}---$\theta_2 > 0$ in Eq.~\eqref{eq:commuting_channels_b} is a \emph{scale} parameter~\cite{jaynes2003probability, linden2014bayesian}, for which the appropriate loss function is logarithmic~\cite{rubio2023quantum}.
Consequently, one must employ a sym\-me\-try-in\-formed loss function of the kind in  Eq.~\eqref{eq:symmetry-msl}. 

A particular realisation, independently enforcing translation invariance for $\theta_1$ and scale invariance for $\theta_2$, is given by
\begin{equation}
    l(\boldsymbol{\tilde{\theta}}, \params) = (\tilde{\theta}_1 - \theta_1)^2 + \log^2\left(\frac{\tilde{\theta}_2}{\theta_2}\right),
    \label{eq:loss-phase-dephasing}
\end{equation}
which mixes two uncertainty types~\cite{overton2024five}.
Here, we assume $L = \id_2$.
This corresponds to using the multiparameter symmetry function
\begin{equation}
    \boldsymbol{f}(\boldsymbol{\theta}) =
    \begin{pmatrix}
        f_1(\boldsymbol{\theta}) \\
        f_2(\boldsymbol{\theta})
    \end{pmatrix}
    =
    \begin{pmatrix}
        \theta_1 \\
        \log(\theta_2)
    \end{pmatrix}.
    \label{eq:symetry-function}
\end{equation}
Crucially, given that Eq.~\eqref{eq:loss-phase-dephasing} preserves the quadratic structure of the square loss \eqref{eq:square_loss}, the bounds derived in this work remain unchanged, except for the definitions of the prior moments,
\begin{subequations}
\begin{align}
\boldsymbol{\mu} &= \int \mathrm{d}\params\, p(\params)\, \boldsymbol{f}(\boldsymbol{\theta}),
\\
\lambda &= \int \mathrm{d}\params\, p(\params)\, \boldsymbol{f}(\boldsymbol{\theta})^{T} L\, \boldsymbol{f}(\boldsymbol{\theta});
\end{align}
\end{subequations}
the first state moments,
\begin{equation}
    \bar{\rho}_i = \int \mathrm{d}\params\, p(\params)\, \rho(\params)\, f_i(\boldsymbol{\theta});
    \label{eq:state-moments-symmetry}
\end{equation}
and the measurement moments,
\begin{subequations}
\begin{align}
    M_i &= \int \mathrm{d}\boldsymbol{\tilde{\theta}} \, M(\boldsymbol{\tilde{\theta}}) \, f_i (\boldsymbol{\tilde{\theta}}),
    \\
    M_{ij} &= \int \mathrm{d}\boldsymbol{\tilde{\theta}} \, M(\boldsymbol{\tilde{\theta}}) \, f_i (\boldsymbol{\tilde{\theta}}) f_j (\boldsymbol{\tilde{\theta}}).
\end{align}
\end{subequations}

Following Ref.~\cite{overton2024five}, Eq.~\eqref{eq:symetry-function} further leads to the ignorance prior \cite{jaynes2003probability}
\begin{equation}
    p(\theta_1,\theta_2) = \frac{1}{\theta_2 W_1 \log(W_2)} \propto \frac{1}{\theta_2}, 
    \label{eq:prior-phase-dephasing}
\end{equation}
with $\theta_1 \in [-W_1/2,\, W_1/2]$ and $\theta_2 \in [1/\sqrt{W_2},\, \sqrt{W_2}]$, 
for which the corresponding \emph{a priori} MSL reads
\begin{equation}
    \mathcal{L}_{\mathrm{prior}} = \lambda = \frac{1}{12}\left[W_1^2 + \log^2(W_2) \right].
\end{equation}
Eq.~\eqref{eq:prior-phase-dephasing} is the product of two independent priors---the first flat, representing maximum ignorance for a location, and the second Jeffreys's prior, representing maximum ignorance for a scale~\cite{jaynes2003probability,linden2014bayesian}.
The range for the dephasing parameter takes such a form when a symmetric, flat range of width $W_2$ is imposed in the log-transformed space.

Given this formulation, the state moments for a single copy can be calculated to have form
\begin{subequations}
\begin{align}
\rho_0 &= \frac{1}{2} [\mathbb{I}_2 + g(W_1, W_2) \, \sigma_x], \\
\bar{\rho}_1 &\propto \sigma_y,  \\
\bar{\rho}_2 &\propto \sigma_x, 
\end{align}
\end{subequations}
with the exact expressions given in App.~\ref{app:phase-dephasing_expressions}.
Although this is a qubit system amenable to further analytic treatment, the resulting expressions are not very informative. 
For our purposes, it suffices to note that the SPM operators arising from Eqs.~\eqref{eq:state-momenta-phase-dephasing} have the following structure:
\begin{subequations}
\begin{align}
    S_1 &\propto \sigma_y, \\
    S_2 &\propto \id_2 + h(W_1, W_2)\,\sigma_x.
\end{align}
\end{subequations}
As can be observed, the SPM operators do not generally commute, thereby guaranteeing the presence of measurement incompatibility at the single-copy level. 
However, the Bayesian weak commutativity condition proposed in Eq.~\eqref{eq:bayes-wc} does hold (as the weak commutativity condition does for the local setting~\cite{vidrighin2014joint}).
This also holds for the many-copy version~\cite{supplemental_code}, which we consider next.

\begin{figure}[!tb]
    \includegraphics[trim={0cm 0cm 0cm 0cm},clip,width=\linewidth]{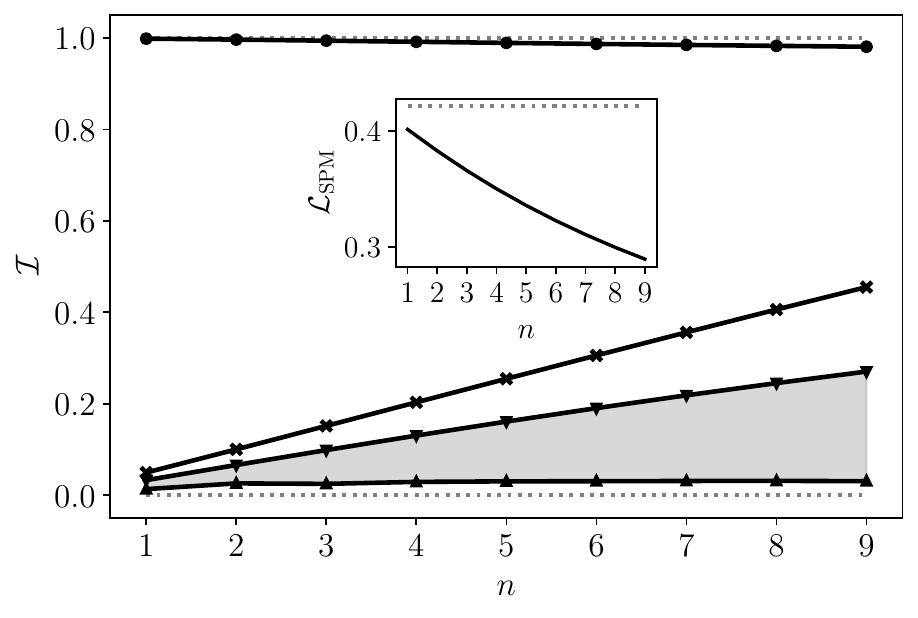}
    \caption{
    Measurement incompatibility bounds from
    the NH bound (upward triangles), the MSL obtained from a tomographic measurement (downward triangles), the standard PGM bound (circles), and the \emph{a priori} MSL (crosses)
    for $n$ copies of the qubit state \eqref{eq:state-phase-diffusion}, with prior widths $W_1 = \pi/2$ and $W_2 = 5$.
    The shaded region represents the range of incompatibility defined in Eq.~\eqref{eq:incomp-range} as a function of $n$.
    The upper and lower bounds on measurement incompatibility Eq.~\eqref{eq:I-between-0-and-1} (dotted) are represented for reference.
    The inset shows the reference MSL, $\mathcal{L}_{\mathsf{SPM}}$,
    which decreases as the number of copies increases,
    as well as the \emph{a priori} MSL,
    which remains constant, leading the corresponding incompatibility to increase.
    The incompatibility associated with the standard PGM bound (circles), is, as in Sec.~\ref{sec:imaging}, is substantially larger than its \emph{a~priori} MSL counterpart and therefore again trivial.
    }
    \label{fig:phase_dephasing}
\end{figure}

Let $\rho^{\otimes n}(\params)$ be the \( n\)-copy state.
Due to the prior, the corresponding \( n \)-copy average state and first moments, i.e., \( \int \mathrm{d} \params \, p(\params) \, \rho^{\otimes n}(\params) \) and \( \int \mathrm{d} \params \, p(\params) \, \rho^{\otimes n}(\params) \, \theta_j \),
are not directly tensor powers of the single-copy \( \rho_0 \) and \( \bar{\rho}_j \) in Eqs.~\eqref{eq:state-momenta-phase-dephasing}.

In lieu of the PGM enhanced by PM estimator bound (due to the high dimension of the \( n \)-copy state), we consider a local qubit tomography---where each copy is measured at random in the basis of one of the three Pauli operators.
The outcomes are then processed with the optimal PM estimator, such that the resulting MSL lower bounds \( \mathcal{L}_{\mathrm{prior}} \), which is again the tighter bound than \( \mathcal{L}_{\mathrm{PGM}} \).

In Fig.~\ref{fig:phase_dephasing} we present numerical results on measurement incompatibility, by evaluating upper and lower bounds as a function of the number of copies, for prior widths $W_1 = \pi/2$ and $W_2 = 5$.
The main plot shows that the incompatibility window grows with the number of copies, with the incompatibility NH lower bound increasing slightly,
and the tomography upper bound increasing more rapidly.
The latter result may be expected, since tomography corresponds to measuring each copy separately and thus cannot take advantage of collective effects.

Interestingly, the overall behaviour appears to be qualitatively different from local QET, where the initially large incompatibility decreases by measuring multiple copies collectively~\cite{vidrighin2014joint,roccia2018entangling}.
In the asymptotic limit we would expect that decrease to manifest, with some eventual reduction in incompatibility.
This is reinforced by the true incompatibility appearing to be at the lower end, with the NH bound being tight for $n=1$ following from App.~\ref{app:NH_tight_qubit_d2},
and appearing attainable numerically by the methods of Sec.~\ref{sec:seesaw-numerics} with \( \abs{\mathcal{I}_{\mathrm{NH}} - \mathcal{I}_{\mathrm{SSW}}} \leq 3{\times}10^{-3} \)
for $n \leq 4$~\cite{supplemental_code}.

Finally, we note that this growth in incompatibility, which may not represent the asymptotic behaviour, does come as the overall estimation precision increases, i.e., $\mathcal{L}_{\mathrm{min}}$ and its upper and lower bounds themselves decrease in the number of copies; see the plot of $\mathcal{L}_{\mathrm{SPM}}$ versus number of copies in the inset.

\subsection{Qubit planar tomography}\label{sec:qubit-tomo}

We conclude by analysing a configuration that, unlike the previous two examples, displays an \emph{a priori} MSL larger than twice the SPM bound---the quantity leading to the upper bound on measurement incompatibility in Eq.~\eqref{eq:I-between-0-and-1}. 
Before proceeding, we note that the purpose of this application is to illustrate specific mathematical properties of the theoretical bounds introduced, rather than to address a problem of physical relevance. For simplicity, we shall therefore use the MSL as a loss quantifier, notwithstanding the symmetry arguments discussed in Secs.~\ref{subsec:estimation-performance} and \ref{sec:phase-dephasing}.

Consider the following family of single-qubit states:
\begin{equation}
\rho(\boldsymbol{\theta}) = \frac{1}{2} \left( \mathbb{I}_2 + \theta_1 \sigma_x + \theta_2 \sigma_y \right),
\end{equation}
where the unknown parameters satisfy the constraint $\theta_1^2 + \theta_2^2 \leq 1$ to ensure physicality.
This setting corresponds to states lying on the plane $z = 0$ of the Bloch sphere, hence the term \emph{planar tomography}.
We assume independent, but not identical, prior distributions for the two parameters, i.e., \( p(\boldsymbol{\theta}) = p_1(\theta_1)\, p_2(\theta_2) \), where \( p_1(\theta_1) \) and \( p_2(\theta_2) \) are supported on \([-W_1, W_1]\) and \([-W_2, W_2]\), respectively.
Overall, we have a bivariate distribution over a rectangle contained inside the Bloch circle in the $x-y$ plane, ensured by the condition $W_1^2 + W_2^2 \leq 1$.
Finally, we assume that the prior for each parameter has zero mean and a finite second moment:
\begin{equation}
    \int \! \mathrm{d} \theta_i \, p_i(\theta_i) \, \theta_i = 0,
    \quad 
    v_i \coloneq \int \! \mathrm{d} \theta_i \, p_i(\theta_i) \, \theta_i^2 \leq \infty.
\end{equation}

Under these assumptions, and choosing $L=\id_2$, the \emph{a priori} MSL is
\begin{equation}
    \mathcal{L}_{\mathrm{prior}} = \lambda = v_1 + v_2 \, ,
\end{equation}
while the zeroth and first state moments are
\begin{equation}
    \rho_0 = \frac{\id}{2}, \quad \bar{\rho}_i = \frac{v_i}{2} \sigma_{i} ,
\end{equation}
where for simplicity we label Pauli matrices as $\sigma_1 \equiv \sigma_x$ and $\sigma_2 \equiv \sigma_y$.
Given that $\rho_0$ is proportional to the identity, the SPM operators are trivially obtained:
\begin{equation}
    S_i = 2 \bar{\rho_i} = v_i \sigma_i.
\end{equation}
Since they do not commute; this estimation problem is impacted by measurement incompatibility.
However, the Bayesian weak commutativity condition proposed in Eq.~\eqref{eq:bayes-wc} does again hold.

The SPM lower bound on the MSL can next be calculated as
\begin{equation}
\begin{split}
    \mathcal{L}_{\mathrm{SPM}}
    &= \lambda - \sum_i \Tr\!\left[\rho_0 S_i^2\right] \\
    &= \lambda - \frac{1}{2}\sum_i v_i^2 \Tr\!\left[\sigma_i^2\right] \\
    &= v_1 + v_2 - \left(v_1^2 + v_2^2\right).
\end{split}
\end{equation}
The NH bound will be calculated numerically, as before.

Moving now to the upper bounds, in this scenario we have $\mathcal{L}_{\mathrm{PGM}} = 2 \mathcal{L}_{\mathrm{SPM}}$, since $\rho_0$ commutes with the first state moments, and the symmetric and square-root inner products are therefore identical.
Unlike in the two previous applications, here we find that the upper bound $2 \mathcal{L}_{\mathrm{SPM}}$ is \emph{not} trivial when
\begin{equation}
    \label{eq:ineq-nontrivial-1qubit-example}
    \mathcal{L}_{\mathrm{prior}} > 2 \mathcal{L}_{\mathrm{SPM}} \iff 2 \left( v_1^2 + v_2^2 \right) > v_1 + v_2.
\end{equation}
Nevertheless, in order to have a tighter upper bound, we also calculate the PGM with the optimal estimator in Eq.~\eqref{eq:MSL_PGMstar}.
Since here such a bound depends on the following quantity:
\begin{equation}
    \Tr\left[ \rho(\params)\, \rho_0^{-\frac{1}{2}} \bar{\rho}_i \rho_0^{-\frac{1}{2}} \right]^2
    = \left( \frac{v_i}{2}\, \theta_i\, \Tr[\sigma_i^2] \right)^2 \\
    = v_i^2 \theta_i^2,
\end{equation}
the integrals over $\theta_i$ with respect to the prior in Eq.~\eqref{eq:MSL_PGMstar} yield a further multiplication by the second moments $v_i$, giving the final expression
\begin{equation}
    \mathcal{L}_{\mathrm{PGM}^*} = v_1 + v_2 - \left( v_1^3 + v_2^3 \right) .
\end{equation}

\begin{figure}[!tb]
    \includegraphics[trim={0cm 0cm 0cm 0cm},clip,width=\linewidth]{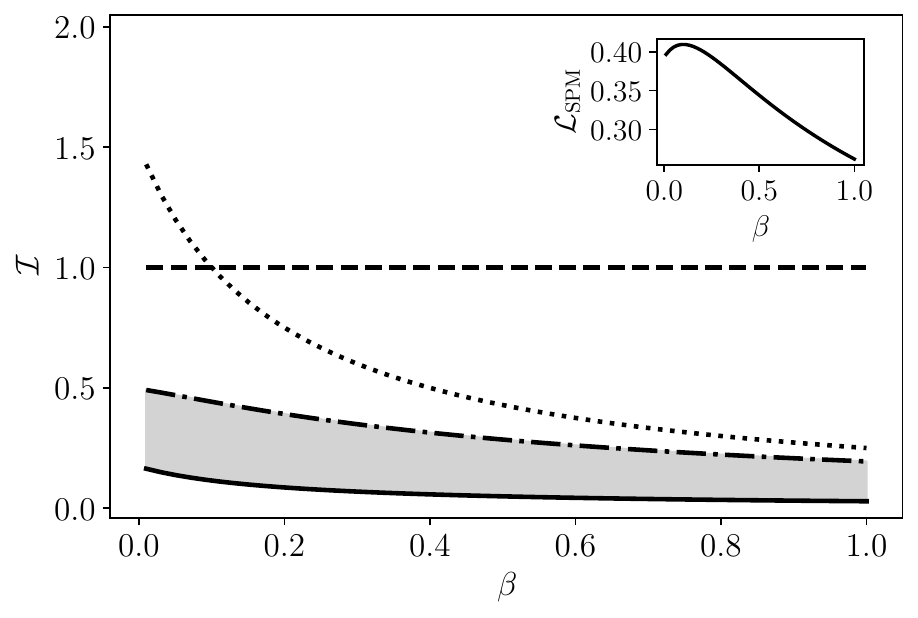}
    \includegraphics[trim={0cm 0cm 0cm 0cm},clip,width=\linewidth]{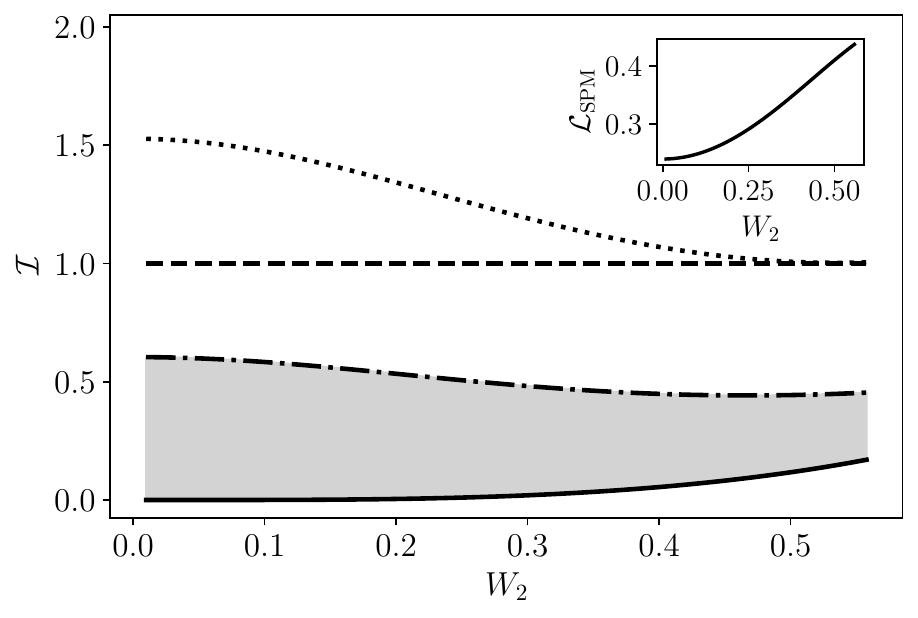}
    \caption{
    Measurement incompatibility bounds from
    the NH bound (solid), the PGM bound enhanced by the PM estimator (dash--dot), the standard PGM bound (dashed), and the \emph{a priori} MSL (dotted).
    Top panel: Measurement incompatibility in qubit planar tomography as a function of the shape parameter $\beta$ of the beta distribution \eqref{eq:beta-prior}, for fixed prior widths $W_1 = 0.85$ and $W_2 = 0.51$.%, for the NH bound (solid), the PGM bound enhanced by the PM estimator (dash--dot), the standard PGM bound (dashed), and the \emph{a priori} MSL (dotted). 
    The shaded region represents the range of incompatibility, as defined in Eq.~\eqref{eq:incomp-range}.
    The inset shows the reference MSL, $\mathcal{L}_{\mathsf{SPM}}$.
    Bottom panel: Analogous configuration but represented as a function of the prior width $W_2$ for the parameter $\theta_2$, for fixed shape parameter $\beta = 0.07$ and prior width $W_1 = 0.83$ for the parameter $\theta_1$.
    Notably, the standard PGM bound is nontrivial in some regimes of these examples, with the prior incompatibility exceeding $1$.}
\label{fig:figure-2}
\end{figure}

We now show a particular choice that satisfies the inequality~\eqref{eq:ineq-nontrivial-1qubit-example}.
Let each independent prior be based on a symmetric beta distribution rescaled and translated such that it is supported on the interval $[-W_i , W_i]$, and having zero mean as required:
\begin{equation}
    p_i(\theta_i) = \frac{1}{2 W_i} h_{\beta}\left( \frac{\theta_i}{2 W_i} + \frac{1}{2} \right), 
    \label{eq:beta-prior}
\end{equation}
where
\begin{equation}
    h_\beta(z) = \frac{(1-z)^{\beta -1} z^{\beta -1}}{B(\beta ,\beta )}, 
\end{equation}
$\beta > 0$ controls the shape,
\begin{equation}
\begin{split}
    B(z_1, z_2) 
    &= \int_0^1 \mathrm{d}t \, t^{z_1 - 1}(1 - t)^{z_2 - 1} \\
    &= \frac{\Gamma(z_1)\,\Gamma(z_2)}{\Gamma(z_1 + z_2)}
\end{split}
\end{equation}
is the beta function, and $\Gamma(z)$ is the gamma function.
For such an \emph{a priori} assignment, the variances are
\begin{equation}
    v_i = \frac{ W_i^2}{ 1 + 2 \beta}.
\end{equation}
Therefore, by appropriately adjusting the parameters $\{W_i\}$ and $\beta$, one can find regions where the inequality~\eqref{eq:ineq-nontrivial-1qubit-example} is satisfied.

The incompatibility bounds corresponding to the MSL upper and lower bounds discussed above are plotted in Fig.~\ref{fig:figure-2}, for various choices of the prior parameters $W_1$, $W_2$ and $\beta$.
The graphical conventions distinguishing curves follow those established in Sec.~\ref{sec:imaging}.
The region of incompatibility values permitted by the bounds is shaded in Fig.~\ref{fig:figure-2}, however, for this class of models the NH bound is exactly attainable (see App.~\ref{app:NH_tight_qubit_d2}), in analogy with local QET~\cite{Nagaoka1991}.

From these plots we observe that, when $\beta$ is sufficiently small---roughly speaking, the prior distribution becomes more concentrated towards the edges---the upper bound $\mathcal{I} \leq \mathcal{I}_{\mathrm{PGM}} = 1$ becomes nontrivial as $\mathcal{I}_{\mathrm{prior}}$ can exceed the value of $1$.
In particular, in the bottom panel of Fig.~\ref{fig:figure-2} we show that for $\beta = 0.07$ and $W_1=0.83$ the inequality $\mathcal{I} \leq 1$ is not trivial for the whole range of valid values for the prior width $0 < W_2 < \sqrt{ 1 - W_1^2 }$, such that $W_2^2 + W_1^2 < 1 $.

\section{Discussion}\label{sec:discussion}

One important aspect that we have not addressed is the relationship between Bayesian QET and the bounds derived in local QET for the frequentist loss in Eq.~\eqref{eq:local-msl}.
In this regard, we recall a conjecture of Ref.~\cite{demkowicz2020multi}: under appropriate regularity conditions, the Bayesian MSL and the frequentist loss would become asymptotically equivalent, that is,
\begin{equation}
    \lim_{n\to\infty} n \mathcal{L}_{\mathrm{min}}^{(n)} = \int \mathrm{d} \params \, p(\params)\, \mathcal{L}_\mathrm{H}(\params),
\end{equation}
where $\mathcal{L}_{\mathrm{min}}^{(n)}$ is the MSL associated with $n$ copies of a given state, $\rho^{\otimes n}(\params)$, and prior $p(\params)$, while $\mathcal{L}_\mathrm{H}(\params)$ is the HCRB of local QET\footnote{Bounds in local QET are most commonly denoted by the letter $C$, which stands for CRB—for example, $C_{\mathrm{H}}(\params)$ for the HCRB.
However, for consistency with the rest of this work we retain the terminology “loss” and the symbol $\mathcal{L}$, since these quantities bound the frequentist loss $\mathcal{L}(\params)$ defined in Eq.~\eqref{eq:local-msl}.}.
This expression would represent the minimum achievable frequentist loss in the asymptotic limit of large $n$.

Since in local QET it holds that~\cite{Tsang2019,Carollo2019} 
\begin{equation}
    2 \mathcal{L}_\mathrm{S}(\params) \geq \mathcal{L}_\mathrm{H}(\params) \geq \mathcal{L}_\mathrm{S}(\params),
\end{equation}
where $\mathcal{L}_\mathrm{S}(\params)$ is the Helstrom (SLD) CRB, our inequality in Eq.~\eqref{eq:2LSPM-upper-bound} is consistent with this asymptotic conjecture.
Notice that, by definition, the Bayesian MSL $\mathcal{L}_{\mathrm{min}}^{(n)}$ is obtained by minimising over POVMs acting collectively on $n$ copies, dubbed ``correlated measurements'' in Ref.~\cite{Hayashi2023a}.

Despite increasing efforts in the experimental implementation of few-copy collective measurements~\cite{Hou2018,roccia2018entangling,Parniak2018beating,Yuan2020,Conlon2023,Mi2025,Yung2025}, scaling them to large $n$ remains challenging in practice.
Following this observation, recent work in local QET has addressed the minimum multiparameter frequentist loss achievable with single-copy measurements and classical feedback, referred to as ``uncorrelated measurements'' in Ref.~\cite{Hayashi2023a}.
This quantity corresponds to a minimisation of the weighted trace of the inverse Fisher information matrix (FIM) over all single-copy POVMs,
\begin{equation}
    \mathcal{L}_{\mathrm{min}}(\params) \coloneq \min_{M(\params)} \Tr[L F(\params)^{-1}],
\end{equation}
where $F(\params)$ is the FIM obtained by applying a POVM to a single copy $\rho(\params)$.
This is often called the ``most informative''~\cite{Nagaoka1989} or ``tight''~\cite{Hayashi2023a} bound.

Unlike the HCRB, $\mathcal{L}_{\mathrm{min}}(\params)$ \emph{can} be larger than $2 \mathcal{L}_\mathrm{S}(\params)$~\cite{Das2024,Hayashi2023a}.
This appears to be in contrast with the Bayesian setting, where the inequality $2 \mathcal{L}_{\mathrm{SPM}} \geq \mathcal{L}_{\mathrm{min}}$ holds at the single-copy level as well.
The resolution of this apparent tension comes from noticing that the $\mathcal{L}_{\mathrm{min}}(\params)$ is still attainable only asymptotically, but restricting each copy to be measured independently, albeit adaptively with classical feedback from previous outcomes.
Therefore, we expect that imposing an analogous restriction in the Bayesian case should lead to a similar behaviour as the number of shots increases.
More concretely, we conjecture that, under suitable regularity conditions
\begin{equation}
    \lim_{n\to\infty} n \mathcal{L}_{\mathrm{min}}^{(n),\mathrm{uncorr}} = \int \! \mathrm{d} \params \, p(\params)  \mathcal{L}_\mathrm{min}(\params) \, ,
\end{equation}
where $\mathcal{L}_{\mathrm{min}}^{(n),\mathrm{uncorr}}$ would be the MSL for the prior $p(\params)$ and family of states $\rho^{\otimes n}(\params)$, but minimised over the uncorrelated measurement strategies defined above, instead of arbitrary collective (correlated) POVMs over all the copies.

We conclude by noting that hybrid figures of merit, incorporating methods from both local and global QET, have been introduced~\cite{Mukhopadhyay2025a,Vasilyev2024}.
However, these approaches are mostly heuristic, and understanding the connections outlined in this discussion could clarify their exact range of applicability.

\section{Conclusions and outlook}\label{sec:conclusions}

QET was initiated in the 1960s and 1970s, but only in recent years have several lines of research converged, leading to a deeper understanding of multiparameter aspects~\cite{demkowicz2020multi,Albarelli2019c,Pezze2025}.
Building on this progress, in this work we have laid the foundations for a theory of Bayesian multiparameter quantum estimation.
We have achieved this by reviewing, streamlining, and contextualising existing results, and by deriving a number of new ones, including: the explicit set of conditions~\eqref{eq:msl-opt-condition} that the optimal POVM must satisfy under the MSL criterion; the upper bound $2 \mathcal{L}_{\mathrm{SPM}}$ on the minimum MSL in the first inequality of Eq.~\eqref{eq:2LSPM-upper-bound}, showing that such a minimum can be at most twice the value of the bound when neglecting measurement incompatibility; the tighter upper bound through the PGM, and a tight bound on both \( \mathcal{L}_{\mathrm{PGM}} \) and \( \mathcal{L}_{\mathrm{prior}} \) using its POVM with the optimal PM estimator; the introduction of general PM operators and their corresponding matrix lower bounds via the monotonicity of inner products, presented in App.~\ref{app:matrix-lower-bounds}; and finally, the demonstration that measurement incompatibility in a single-shot Bayesian setup is substantially affected by prior information, as illustrated through examples in qubit and optical configurations.

This work opens several avenues for future research that we now highlight, both theoretical and experimental.
For example, in multiparameter quantum metrology---the simultaneous estimation of several parameters encoded by a quantum channel---additional precision trade-offs may arise from the choice of probe state and control strategy~\cite{ragy2016compatibility,Gorecki2020,Hayashi2023a,Hu2024a}, a phenomenon referred to as \emph{probe incompatibility}~\cite{Albarelli2022}.
While this aspect has not, to the best of our knowledge, been examined in detail and in full generality within a Bayesian framework, numerical approaches for designing optimal multiparameter Bayesian estimation protocols~\cite{Kaubruegger2023,Bavaresco2024designing,Andre2026} inherently account for probe incompatibility effects.

Still outstanding are conditions that lead to collapse in the hierarchy of lower or upper bounds.
Perhaps the most likely candidate would be the Bayesian weak commutativity condition proposed in Eq.~\eqref{eq:bayes-wc}, whose local QET analogue implies equivalence of the Helstrom (SLD) and Holevo CRBs.
Entirely consistent with the Bayesian weak-commutativity condition implying saturability of the NH bound,
all our examples satisfy Eq.~\eqref{eq:bayes-wc} and the NH bound appears attainable.
Incidentally, the Bayesian Holevo bound also matches the SPM bound numerically in all examples%
\footnote{Supplemental data~\cite{supplemental_code} shows that the Bayesian Holevo bound matches the SPM bound
  with \( \abs{\mathcal{I}_{\mathrm{H}}} \leq 2{\times}10^{-7} \) for $2 \leq d\ \leq 17$ for the example considered in Sec.~\ref{sec:imaging},
  with \( \abs{\mathcal{I}_{\mathrm{H}}} \leq 2{\times}10^{-5} \) for $1 \leq n \leq 6$
  (the Bayesian Holevo bound calculation becomes highly computationally demanding for larger $n$) 
  for the example considered in Sec.~\ref{sec:phase-dephasing},
  and with \( \abs{\mathcal{I}_{\mathrm{H}}} \leq 3{\times}10^{-5} \) for all points shown in Fig.~\ref{fig:figure-2}.},
and---though not saturable and generally less significant in the Bayesian framework---this
may be expected to hold more generally when the Bayesian weak-commutativity condition is satisfied,
given the similarity of the underlying algebra.

Another important problem is addressing in\-com\-pat\-i\-bil\-i\-ty---measurement and other forms thereof---for parameters obeying symmetries other than translation invariance.
The symmetry remapping in Eq.~\eqref{eq:symetry-function} provides a starting point to generalise the results of this study to any location-isomorphic estimation problem, concrete illustrations of which have been discussed both in Sec.~\ref{sec:phase-dephasing} and in Ref.~\cite{overton2024five}.
A rigorous formulation should include the possibility of correlated priors and loss functions coupling different parameters, as well as the application of information geometry~\cite{boeyens2021uninformed}.
Moreover, a further extension to symmetries beyond location-isomorphic metrology is necessary to encompass phase parameters~\cite{holevo2011probabilistic,Demkowicz-Dobrzanski2011}.

In addition, while our analysis has focused on fi\-nite-di\-men\-sion\-al
systems, continuous-variable (CV) platforms---par\-tic\-u\-lar\-ly quantum optical ones~\cite{demkowiczdobrzanski2015quantum,Polino2020,Pirandola2018,Barbieri2022}---play a central role in sensing applications. 
It will therefore be important to rigorously extend the theoretical framework developed here to CV systems.
Given the rapid progress in learning theory for CV states~\cite{Wu2024g,Mele2025,Bittel2025,Fanizza2025}, this direction appears especially promising.
For example, the case of Gaussian states and priors \cite{wang2007} may allow for a tractable analysis while remaining relevant for practical implementations.
In single-parameter Bayesian estimation, several specific results for Gaussian states are already known~\cite{Morelli2020}, and a general framework for multiparameter local QET with Gaussian states---capable of giving incompatibility with collective measurements---has recently been proposed~\cite{Chang2025}.

At a practical level, identifying optimal strategies for even an individual shot in an adaptive, experimental protocol can be computationally difficult.
Even methods reducible to convex problems~\cite{Bavaresco2024designing} are currently far less performant than solving the corresponding NH bound, in turn more demanding than the SPM and PGM lower and upper bounds.
The upper bound $\mathcal{L}_{\mathrm{min}} \leq 2 \mathcal{L}_{\mathrm{SPM}}$ suggests that optimising the SPM affords a good compromise between optimality and tractability, compared to the exact but difficult minimum loss \eqref{eq:msl-min}, with a smaller SPM bound also meaning a smaller upper bound on \( \mathcal{L}_{\mathrm{min}} \) and so the worst-case impact of measurement-incompatibility is smaller in absolute terms.
In practice this sees the pseudo-gain $L^{ij} \Tr[\rho_0 S_i S_j]$---i.e., the second term in the SPM lower bound~\eqref{eq:SPM_lower_bound}; see App.~\ref{app:prior-loss}---become the optimiser of experimental control parameters for individual shots, fulfilling the role of a global alternative to the weighted trace of the inverse quantum FIM. 
Although such an approach will \emph{not} generally lead to full optimality, we anticipate a substantial boost in quantum technological applications, in light of recent results from cold-atom experiments showing two-fold~\cite{glatthard2022optimal} and five-fold~\cite{overton2024five} precision enhancements obtained by employing the “classical” version of this quantifier, that is, the pseudo-gain computed using the PM estimator and a fixed POVM.

Altogether, we expect the Bayesian multiparameter framework for QET introduced here, together with the insight into the impact of measurement incompatibility, to become a fundamental building block for benchmarking both theory and experiment in quantum metrology, guiding the design of practically relevant, information-optimal quantum sensors.

\vspace{10pt}
\emph{Note added.}~During the preparation of the final version of this manuscript for publication, we became aware of the independent preprint by Zhang, Yamagata, and Suzuki~\cite{Zhang2026n}, which develops a systematic theory of Bayesian monotone metrics and contains, among other results, a derivation of the monotone-metric matrix and scalar lower bounds presented in App.~\ref{app:matrix-lower-bounds}.
Revisiting our derivation, we found that an earlier version of App.~\ref{app:matrix-lower-bounds} should have used complex, rather than only real, linear combinations of the parameter moments in order to justify the complex matrix inequality.
We have amended the proof accordingly; the stated bounds are unchanged.
Previous versions also specified an unnecessary property that linear maps $E_{\rho}$ must satisfy; we thank Jun Suzuki for highlighting this.
This property was not relied on in the subsequent proof.

% Acknowledgements
\begin{acknowledgments}
F.~A. acknowledges fruitful discussions with \mbox{Stefano} \mbox{Chessa}, \mbox{Rafał} \mbox{Demkowicz-Dobrzański}, \mbox{Vittorio} \mbox{Giovannetti} and \mbox{Ludovico} \mbox{Lami}.
J.~R. gratefully thanks \mbox{Luis} \mbox{Correa} and \mbox{Edward} \mbox{Gandar} for insightful discussions.
F.~A. acknowledges partial financial support from Marie Skłodowska-Curie Action EUHORIZON-MSCA-2021PF-01 (project \mbox{QECANM}, Grant No.~101068347).
D.~B. acknowledges financial support from the European Union's Horizon Europe research and innovation program under the EPIQUE Project (Grant Agreement No.~101135288).
J. R. acknowledges financial support from the Surrey Future Fellowship (University of Surrey), during which this project was developed and first released, and partial support from the UKRI-JST project ``Quantum Control \& Sensing: Enhancing high-precision quantum sensing in noisy environments via optimised control'', during which this work was brought to publication.

Code and data generated for this paper are openly available~\cite{supplemental_code}.
An open-source python package for evaluating the bounds discussed is also available on GitHub~\cite{multipar2025code}.
\end{acknowledgments}

\section*{Author contributions}
F.~A. and J.~R. conceived the project, developed the theoretical framework and performed analytical calculations.
D.~B. carried out numerical simulations and also contributed to the theoretical framework and analytical calculations.
All authors discussed the results and wrote the manuscript.
The derivations in Appendix \ref{app:NH_tight_qubit_d2} and the example in Sec.~\ref{sec:qubit-tomo} were obtained in a conversational style with a large language model (ChatGPT 5).
This tool was also employed to assist with text editing and code optimisation.
All scientific content and final text were reviewed and approved by the authors, who take full responsibility for the work.

% References
%\nocite{apsrev42Control}
\bibliography{refs_arXiv}

% \printbibliography

% Appendices
\onecolumngrid
\appendix

\section{Prior loss and precision gain in Bayesian quantum estimation}\label{app:prior-loss}

Suppose we make a random guess of the unknown parameters based solely on the prior probability.  
Let us denote the corresponding vector of estimates by $\boldsymbol{\tilde{\theta}}$, noting the absence of any outcome dependence.
The MSL in this case is given by
\begin{equation}
    \mathcal{L}_{p} (\boldsymbol{\tilde{\theta}}) 
    \coloneq 
    \int \mathrm{d}\boldsymbol{\theta}\, p(\boldsymbol{\theta})\, 
    (\boldsymbol{\tilde{\theta}} - \boldsymbol{\theta})^{T}
    L\, 
    (\boldsymbol{\tilde{\theta}} - \boldsymbol{\theta}),
    \label{eq:prior-loss-general}
\end{equation}
which we may refer to as the \emph{prior loss}.
Alternatively, one may obtain Eq.~\eqref{eq:prior-loss-general} by inserting a trivial POVM consisting of a single element, the identity operator, into the MSL in Eq.~\eqref{eq:msl_compact}.

It is straightforward to verify that the estimator minimising Eq.~\eqref{eq:prior-loss-general} is the prior mean,
\begin{equation}
    \boldsymbol{\mu} 
    \coloneq 
    \int \mathrm{d}\boldsymbol{\theta}\, p(\boldsymbol{\theta})\, \boldsymbol{\theta},
    \label{eq:prior-mean}
\end{equation}
which leads to the minimum prior loss
\begin{equation}
\begin{aligned}
    \mathcal{L}_{\mathrm{prior}} 
    \coloneq& \,\, \mathcal{L}_{p} (\boldsymbol{\mu}) 
    \\
    =& \int \mathrm{d}\boldsymbol{\theta}\, p(\boldsymbol{\theta})\, 
    \boldsymbol{\theta}^{\top} 
    L\, \boldsymbol{\theta} - \boldsymbol{\mu}^T\, L\,
    \boldsymbol{\mu} 
    \\
    =& \,\lambda - \boldsymbol{\mu}^T\,L\,
    \boldsymbol{\mu},
\end{aligned}
\end{equation}
where $\lambda$ was already defined in Eq.~\eqref{eq:prior-second-moment}.
Such an optimal prior loss is nothing but $\tr L \Sigma_p $, where $\Sigma_p$ is the covariance matrix of the multivariate prior probability $p(\params)$ and we have introduced a lowercase trace of $d{\times}d$ matrices, $\tr$, distinct from the uppercase trace of operators acting on the physical Hilbert space of the system.

One may further show that $\mu_i = \Tr[\rho_0 S_i]$, where $\rho_0$ and $S_i$ are defined in Sec.~\ref{sec:opt-quantum-strategies}.
This, together with Eq.~\eqref{eq:prior-mean}, allows us to rewrite the SPM bound in Eq.~\eqref{eq:SPM_lower_bound} as 
\begin{equation}
    \mathcal{L}_{\mathrm{SPM}} = \mathcal{L}_{\mathrm{prior}} - \mathcal{G}_{\mathrm{SPM}},
\end{equation}
where 
\begin{equation}
    \label{eq:scalar-precision-gain}
    \mathcal{G}_{\mathrm{SPM}} \coloneq L^{ij} (\Tr[\rho_0 S_j S_i] - \Tr[\rho_0 S_j]\,\Tr[\rho_0 S_i]).
\end{equation}
Written in this form, the SPM bound can be interpreted as the result of reducing the prior loss by the amount of precision gained through individually optimal measurements, $\mathcal{G}_{\mathrm{SPM}}$.

Recent literature has referred to the quantity $L^{ij} \Tr[\rho_0 S_j S_i]$ as an alternative definition of precision gain when the SPM bound is used as an optimiser for control parameters~\cite{overton2024five, rubio2025on}. 
From a computational standpoint, this definition is often more convenient, and it has likewise been adopted in our code~\cite{multipar2025code}. 
Accordingly, we refer to this quantity as the \emph{pseudo-gain}, following the discussion above. 
Naturally, the gain and pseudo-gain coincide whenever the vector prior mean $\boldsymbol{\mu}$ vanishes.

\section{Optimality conditions for the MSL}\label{app:fundamental-eqs}

Let $M_{\mathrm{opt}}(\boldsymbol{\tilde{\theta}})$ denote the optimal POVM for a Bayesian quantum multiparameter estimation problem.  
Following Holevo~\cite{Holevo1973,Holevo1973b,kholevo1978investigations} and Helstrom \cite{helstrom1974noncommuting, helstrom1976quantuum}, a necessary and sufficient condition for such a POVM to exist is the existence of a Hermitian operator
\begin{equation}
    \Upsilon = \int \mathrm{d}\boldsymbol{\tilde{\theta}}\,
    M_{\mathrm{opt}}(\boldsymbol{\tilde{\theta}})\,
    \varrho(\boldsymbol{\tilde{\theta}}) 
    = \int \mathrm{d}\boldsymbol{\tilde{\theta}}\,
    \varrho(\boldsymbol{\tilde{\theta}})\,
    M_{\mathrm{opt}}(\boldsymbol{\tilde{\theta}}),
    \label{eq:holevo-helstrom-condition-1}
\end{equation}
with
\begin{equation}
    \varrho(\boldsymbol{\tilde{\theta}}) \coloneq 
    \int \mathrm{d}\params\, p(\params)\,\rho(\params)\,
    l(\boldsymbol{\tilde{\theta}}, \boldsymbol{\theta}),
    \label{eq:quantum-risk}
\end{equation}
that satisfies
\begin{equation}
    \varrho(\boldsymbol{\tilde{\theta}}) - \Upsilon \succeq 0
    \label{eq:holevo-helstrom-condition-2}
\end{equation}
for all $\boldsymbol{\tilde{\theta}}$.  
The minimum loss is then given by $\mathcal{L}_{\mathrm{min}} = \Tr(\Upsilon)$.  
A discrete version of this optimality condition exists for hypothesis testing \cite{Yuen1970,Yuen1975}, and a generalisation to convex functions has recently been proposed \cite{Coutts2018}.

We now derive the explicit form of these conditions for the MSL.  
Using the notation introduced in Sec.~\ref{sec:opt-quantum-strategies}, Eq.~\eqref{eq:quantum-risk} becomes
\begin{equation}
    \varrho(\boldsymbol{\tilde{\theta}}) 
    = \Lambda + L^{ij} 
    \bigl(\tilde{\theta}_j \tilde{\theta}_i \rho_0 - 2 \tilde{\theta}_j \bar{\rho}_i \bigr),
    \label{eq:quantum-risk-msl}
\end{equation}
where $\Tr(\Lambda) = \lambda$.  
Accordingly, the operator in Eq.~\eqref{eq:holevo-helstrom-condition-1} takes the form
\begin{equation}
    \Upsilon 
    = \Lambda + L^{ij}\bigl(
        M_{\mathrm{opt}, ji} \rho_0 
        -
        2 M_{\mathrm{opt}, j} \bar{\rho}_i
      \bigr)
    = \Lambda + L^{ij}\bigl(
        \rho_0 M_{\mathrm{opt}, ji} - 2 \bar{\rho}_j M_{\mathrm{opt}, i}
      \bigr).
\end{equation}
By adding both expressions for $\Upsilon$ and dividing by two, we obtain
\begin{equation}
    \Upsilon = \Lambda + \tfrac{1}{2} L^{ij}(
        M_{\mathrm{opt}, ji}\rho_0 + \rho_0 M_{\mathrm{opt}, ji})
        - M_{\mathrm{opt}, j} \bar{\rho}_i - \bar{\rho}_j M_{\mathrm{opt}, i}.
    \label{eq:hermitian-op-msl}
\end{equation}
Finally, inserting Eqs.~\eqref{eq:quantum-risk-msl} and \eqref{eq:hermitian-op-msl} into Eq.~\eqref{eq:holevo-helstrom-condition-2}, we obtain the conditions that any POVM optimal under the MSL criterion must satisfy:
\begin{equation}
    L^{ij} [
        (\tilde{\theta}_j \tilde{\theta}_i \mathbb{I} - M_{\mathrm{opt}, ji}) \rho_0
        + \rho_0 (\tilde{\theta}_j \tilde{\theta}_i \mathbb{I} - M_{\mathrm{opt}, ji})
        - 2 (\tilde{\theta}_j \mathbb{I} - M_{\mathrm{opt}, j}) \bar{\rho}_i
        - 2 \bar{\rho}_j (\tilde{\theta}_i \mathbb{I} - M_{\mathrm{opt}, i})
    ] \succeq 0,
    \label{eq:msl-opt-condition-app}
\end{equation}
for all $\boldsymbol{\tilde{\theta}}$, as stated in the main text.

\section{Tightness of the NH bound for two parameters on a single qubit}
\label{app:NH_tight_qubit_d2}

In the main text we have shown that the Bayesian NH bound is obtained by relaxing the requirement that the first- and second-moment operators $\{M_i\}$ and $\{M_{ij}\}$ originate from an underlying POVM $M(x)$, exploiting the operator-valued moment-matrix inequality~\eqref{eq:lemma1-Suzuki}.
In general, this relaxation can be loose.
However, in this appendix we show that, for two parameters on a single qubit, the relaxation is instead exact, and it coincides with an attainability result proved by Nagaoka for $2 {\times }2$ matrices.
Thus, this appendix shows that the reason for the tightness of Nagaoka's two-parameter bound in local estimation naturally carries over to the Bayesian domain.

For $d=2$ and the identity weight matrix $L= \id_2$ (without loss of generality; see footnote 4 in the main text), the optimisation over the second-moment operators $M_{11},M_{12},M_{22}$ can be carried out analytically, leading to Eq.~\eqref{eq:Nagaoka2par}, which we reproduce here for convenience:
\begin{equation}
\mathcal{L}_{\mathrm{NH}}
=\lambda+\min_{M_1,M_2}\left\{
\Tr\!\big[\rho_0(M_1^2+M_2^2)\big]
+\big\|\sqrt{\rho_0}[M_1,M_2]\sqrt{\rho_0}\big\|_1
-2\Tr\!\big[\bar\rho_1M_1+\bar\rho_2M_2\big]
\right\}.
\label{eq:app_A1}
\end{equation}
Here, $M_1,M_2$ are the operator-valued first moments, defined in Eq.~\eqref{eq:measurement-moments} in the main text, and $\|\cdot\|_1$ is the trace norm.
The only source of looseness is that, in the derivation of $\mathcal{L}_{\mathrm{NH}}$, the pair $(M_1,M_2)$ is treated as a free pair of Hermitian operators, constrained only indirectly through Eq.~\eqref{eq:lemma1-Suzuki}, without explicitly enforcing that it comes from a common POVM $M(x)$.

Nagaoka proved the following result, for which we adapt the notation to the one used in this paper:
\begin{externaltheorem}[{Ref.~\cite[Theorem~2]{Nagaoka1991}}]
Fix a pair of Hermitian operators $M_1,M_2$ acting on a two-dimensional Hilbert space, and consider the set of all POVMs $M(x_1,x_2)$ on $\mathbb{R}^2$ whose first moments match this pair:
\begin{equation}
\int \! \mathrm{d} x_1 \mathrm{d} x_2 \, M(x_1,x_2) \,  x_i =M_i,\qquad i=1,2.
\label{eq:app_A2}
\end{equation}
Among these POVMs, define the minimal achievable ``quadratic'' second-moment cost $\sum_i\Tr[\rho_0 M_{ii}]$ to be
\begin{equation}
C^\star(M_1,M_2)
:=\inf_{M(x_1,x_2)\ \text{s.t.\ \eqref{eq:app_A2}}}
\Tr\!\left[\rho_0\int \! \mathrm{d} x_1 \mathrm{d} x_2 \, (x_1^2+x_2^2) \,M(x_1, x_2)\right].
\label{eq:app_A3}
\end{equation}
Then, for a qubit system, this infimum is attainable and admits the closed form
\begin{equation}
C^\star(M_1,M_2)
=\Tr\!\big[\rho_0(M_1^2+M_2^2)\big]
+\big\|\sqrt{\rho_0}[M_1,M_2]\sqrt{\rho_0}\big\|_1 \,.
\label{eq:app_A4}
\end{equation}
\end{externaltheorem}
This means that, for any $M_1$, $M_2$, an MSL
\begin{equation}
\mathcal{L}
=\lambda+
\Tr\!\big[\rho_0(M_1^2+M_2^2)\big]
+\big\|\sqrt{\rho_0}[M_1,M_2]\sqrt{\rho_0}\big\|_1
-2\Tr\!\big[\bar\rho_1M_1+\bar\rho_2M_2\big]
\label{eq:app_A5}
\end{equation}
is attainable,
which entails attainability of the Bayesian NH lower bound Eq.~\eqref{eq:app_A1}.
Nagaoka also provided explicitly an optimal POVM attaining the equality above.
Notice that this result is proven under the assumption $\rho_0 \succ 0$, which, for qubit systems, holds in essentially all nontrivial cases. 
The only exception arises when the prior is a Dirac delta, in which case the estimation problem becomes trivial.

The Bayesian NH bound is obtained by optimising over $\{M_i\}$ and $\{M_{ij}\}$ subject to the moment-matrix inequality~\eqref{eq:lemma1-Suzuki}, instead of optimising over POVMs directly.
For two parameters, Eq.~\eqref{eq:Nagaoka2par}
(restated as Eq.~\eqref{eq:app_A1}) already encodes the best possible contribution of second moments compatible with a given pair $(M_1,M_2)$, namely Eq.~\eqref{eq:app_A4}.
Therefore, in the qubit case for every candidate pair $(M_1,M_2)$, there exists a POVM whose second moments attain the value
$\Tr[\rho_0(M_1^2+M_2^2)] + \|\sqrt{\rho_0}[M_1,M_2]\sqrt{\rho_0}\|_1$; hence the minimisation in Eq.~\eqref{eq:app_A1} is not merely a relaxation: it coincides with the true minimum MSL over single-qubit POVMs.
Operationally, one may view Eq.~\eqref{eq:app_A1} as first optimising over the ``effective first-moment operators'' $M_1,M_2$, and then invoking Nagaoka's construction to realise the corresponding optimal second moments by a concrete measurement on the qubit.

\subsection{Explicit optimal POVM attaining the two parameter single qubit NH bound}\label{app:NH_tight_qubit_d2:povm}
For completeness, we briefly give the details of an explicit POVM attaining Eq.~\eqref{eq:app_A4} for any fixed Hermitian pair $(M_1,M_2)$ on a qubit with a full-rank $\rho_0$.
We assume that $M_1$ and $M_2$ are linearly independent and non-commuting; otherwise, the problem is trivial, as they would share a common eigenbasis and the SPM bound would therefore be tight.
Let $x := \Tr[\rho_0 M_1]$ and $y := \Tr[\rho_0 M_2]$ in order to define the centred operators $X := M_1 - x \id_2$ and $Y := M_2 - y \id_2$, so that $\Tr[\rho_0 X]=\Tr[\rho_0 Y]=0$.
Introduce an inner product induced by the state $\rho_0$ (see the next section for more details on such inner products)
\begin{equation}
\label{eq:SLD-inner-qubit}
\langle A,B\rangle_{\rho_0}^{\mathrm{S},(e)} := \frac{1}{2} \Tr \left[ \rho_0 \left( A B + B A \right) \right],
\end{equation}
and let $\mathcal{W}:=\mathrm{span}_{\mathbb{R}}\{X,Y\}$, i.e., the two-dimensional subspace of Hermitian operators with zero first moments spanned by $X$ and $Y$.
Define a linear map $\mathsf{A}:\mathcal{W}\to \mathcal{W}$ as
\begin{equation}
\mathsf{A}(W) := \langle W,X\rangle_{\rho_0}^{\mathrm{S},(e)}\,X \;+\; \langle W,Y\rangle_{\rho_0}^{\mathrm{S},(e)}\,Y .
\end{equation}
This map is symmetric and positive definite with respect to $\langle\cdot,\cdot\rangle_{\rho_0}^{\mathrm{S},(e)}$, hence it has an orthonormal eigenbasis of Hermitian operators $\{U_1,U_2\}\subset \mathcal{W}$ with eigenvalues $\nu_1,\nu_2>0$ satisfying
$\mathsf{A}(U_j)=\nu_j U_j$ and $\langle U_j,U_k\rangle_{\rho_0}^{\mathrm{S},(e)}=\delta_{jk}$.

Now, the optimal POVM can be obtained as a random projective measurement on the eigenbasis of $U_1$ and $U_2$.
Explicitly, diagonalise each $U_j$ as $U_j=\sum_{k=1}^2 u_{jk} E_{jk}$, where $u_{jk}\in\mathbb{R}$ are eigenvalues and $E_{jk}$ are rank-one projectors with $\sum_{k=1}^2 E_{jk}=\id_2$.
Define
\begin{equation}
p_j := \frac{\sqrt{\nu_j}}{\sqrt{\nu_1}+\sqrt{\nu_2}}, \qquad 
M(j,k) := p_j\, E_{jk},
\end{equation}
so that $\{M(j,k)\}_{j,k\in\{1,2\}}$ is a 4-outcome POVM satisfying $\sum_{j,k}M(j,k)=\id_2$ .
Now, introducing the coefficients
\begin{equation}
x_{jk} := p_j^{-1}\,\langle X,U_j\rangle_{\rho_0}^{\mathrm{S},(e)}\,u_{jk} + x ,\qquad
y_{jk} := p_j^{-1}\,\langle Y,U_j\rangle_{\rho_0}^{\mathrm{S},(e)}\,u_{jk} + y,
\end{equation}
the first-moment constraints hold as follows: 
$\sum_{j,k} x_{jk} M(j,k) = M_1$ and $\sum_{j,k} y_{jk} M(j,k) = M_2$.
It can be checked that the second-moment cost for this POVM and coefficients $\{x_{jk}, y_{jk}\}$ attains the minimum in Eq.~\eqref{eq:app_A3}:
\begin{equation}
\Tr\left[ \rho_0 \sum_{jk} ( x_{jk}^2 + y_{jk}^2  ) M(j,k) \right] =x^2+y^2+(\sqrt{\nu_1} + \sqrt{\nu_2} )^2 = x^2+y^2 +  
\Tr[ \rho_0  (X^2 + Y^2 )  ] + \left\Vert \sqrt{\rho_0}  [X, Y ] \sqrt{\rho_0} \right\Vert_1;
\label{eq:final_Nagaoka_optimalPOVM}
\end{equation}
this matches Eq.~\eqref{eq:app_A4}, since
$x^2+y^2 +
\Tr[ \rho_0  (X^2 + Y^2 )  ] = \Tr[ \rho_0  (M_1^2 + M_2^2 )  ] $,
and $ [X, Y ] = [M_1,M_2]$.

\subsection{Closed form of the qubit NH bound}
\label{app:NH_qubit_closed_form}

We now show that, for two parameters and a single qubit, the remaining minimisation in Eq.~\eqref{eq:app_A1} also admits a closed form.
We only prove the result for $L=\id_2$, as the bounds for other loss matrices can be obtained simply by reparametrization of the first moment operators (see footnote 4 in the main text).

Let $S_i$ be the SPM operators for the qubit model, i.e., $\bar{\rho}_i = E_{\rho_0,\mathrm{S}}[S_i] =\frac{S_i\rho_0+\rho_0S_i}{2}$ with $i=1,2$, and also introduce their zero-mean version $ \Delta S_i:=S_i-\mu_i\id_2 $, where $\mu_i = \Tr[\rho_0S_i]$ are the prior mean values in Eq.~\eqref{eq:prior-mean}.
Introducing the real $2\times2$ Gram matrix of the zero-mean SPM operators
\begin{equation}
    \mathcal{H}{ij}:=
    \frac12\Tr\!\left[\rho_0\{\Delta S_i,\Delta S_j\}\right]
    =\Re\Tr[\rho_0\Delta S_i\Delta S_j] = \langle \Delta S_i,\Delta S_j \rangle_{\rho_0}^{\mathrm{S},(e)} ,
    \label{eq:Gamma_qubit_NH}
\end{equation}
which corresponds to the matrix-valued version of the precision gain in Eq.~\eqref{eq:scalar-precision-gain}, the two-parameter qubit Bayesian NH bound is equivalent to
\begin{equation}
    \mathcal{L}_{\mathrm{NH}}
    = 
    \lambda-\mu_1^2-\mu_2^2-\lambda_{\max}(\mathcal{H}) =  \mathcal L_{\mathrm{prior}} -\lambda_{\max}(\mathcal{H}) = \mathcal{L}_{\mathrm{SPM}}
    +\lambda_{\min}(\mathcal{H}) , 
    \label{eq:NH_qubit_closed_1}
\end{equation}
where $\lambda_{\max}(\mathcal{H})$ and $\lambda_{\min}(\mathcal{H})$ denote the largest and smallest eigenvalues of of $\mathcal{H}$, respectively.
The last equality holds because for $L=\id_2$ the SPM bound can be written as $\mathcal L_{\mathrm{SPM}}  = \lambda-\sum_{i=1}^2\Tr[\bar\rho_iS_i] = \lambda-\mu_1^2-\mu_2^2-\tr\mathcal{H}$ and $\tr \mathcal{H} = \lambda_{\max}(\mathcal{H}) + \lambda_{\min}(\mathcal{H}) $.
The quantity $\lambda_{\min}(\mathcal{H})$ can be written explicitly in terms of a commutator between the SPM operators, yielding
\begin{equation}
    \mathcal{L}_{\mathrm{NH}}
    =
    \mathcal{L}_{\mathrm{SPM}}
    +\frac{1}{2}\left[
        \tr\mathcal{H}
        -
        \sqrt{(\tr\mathcal{H})^2
        -\left\|
            \sqrt{\rho_0}[S_1,S_2]\sqrt{\rho_0}
        \right\|_1^2}
    \right].
    \label{eq:NH_qubit_closed_comm}
\end{equation}

\begin{proof}
Starting from Eq.~\eqref{eq:app_A1}, one can write any Hermitian first-moment operator $M_i$ in terms of its zero-mean version $X_i$ as
\begin{equation}
    M_i=m_i\id_2+X_i,\qquad \Tr[\rho_0X_i]=0 .
\end{equation}
From the SPM definition we have
\begin{equation}
    \Tr[\bar{\rho}_iM_i]
    =
    \frac12\Tr[\rho_0(S_iM_i+M_iS_i)]
    =
    \mu_i m_i+\Re\Tr[\rho_0\Delta S_iX_i],
\end{equation}
where the mixed terms vanish because both $X_i$ and $\Delta S_i$ have zero mean with respect to $\rho_0$.
Since the commutator term is invariant under shifts by multiples of the identity, the optimisation in Eq.~\eqref{eq:app_A1} separates into a scalar part and a part involving the zero-mean operators.
\begin{align}
& \min_{M_1,M_2}\left\{
\Tr\!\big[\rho_0(M_1^2+M_2^2)\big]
+\big\|\sqrt{\rho_0}[M_1,M_2]\sqrt{\rho_0}\big\|_1
-2\Tr\!\big[\bar\rho_1M_1+\bar\rho_2M_2\big]
\right\} \\ 
& = \min_{m_1,m_2} \sum_{i=1}^2 \left(  m_i^2-2\mu_i m_i \right) +  
\min_{X_1,X_2} \left\{  \sum_{i=1}^2  \left( \Tr \left[ \rho_0 X_i^2 \right] - 2 \Re\Tr[\rho_0\Delta S_i X_i ] \right)
 + \big\|\sqrt{\rho_0}[X_1,X_2]\sqrt{\rho_0}\big\|_1 \right\} \\ 
& = -\mu_1^2-\mu_2^2  +  
\min_{X_1,X_2} \left\{  \sum_{i=1}^2 \left(  \Tr \left[ \rho_0 X_i^2 \right] - 2 \Re\Tr[\rho_0\Delta S_i X_i ] \right)
 + \big\|\sqrt{\rho_0}[X_1,X_2]\sqrt{\rho_0}\big\|_1 \right\} ,
\label{eq:NHqubitexplicit-zeromean-objective}
\end{align}
where the scalar part has been explicitly minimised by $m_i=\mu_i$.

Thus, what remains is a minimisation over the zero-mean Hermitian operators $X_1,X_2$ which belong to the real vector space $\mathcal{V}:=\{X=X^\dagger:\Tr[\rho_0X]=0\}$.
We employ the symmetric inner product $\langle A,B\rangle_{\rho_0}^{\mathrm{S},(e)}$ defined in Eq.~\eqref{eq:SLD-inner-qubit} and introduce the Gram matrix of the operators $X_1, X_2$ (denoted collectively as $\mathsf{X}$): 
\begin{equation}
\Phi[\mathsf{X}]_{ij} = \langle X_i,X_j\rangle_{\rho_0}^{\mathrm{S},(e)}.
\end{equation}
This Gram matrix can be used to express part of the objective function in Eq.~\eqref{eq:NHqubitexplicit-zeromean-objective}.
First, we have
\begin{equation}
   \Tr \left[ \rho_0 \left( X_1^2 + X_2^2 \right) \right] = \langle X_1,X_1\rangle_{\rho_0}^{\mathrm{S},(e)}
    +
    \langle X_2,X_2\rangle_{\rho_0}^{\mathrm{S},(e)}
    =
    \tr \Phi[\mathsf X].
\end{equation}
Second, for a single qubit, one can check explicitly the following identity by Nagaoka~\cite[Lemma 8]{Nagaoka1991}
\begin{equation}
    \label{eq:Nagaoka-identity}
    \left\|
        \sqrt{\rho_0}[X_1,X_2]\sqrt{\rho_0}
    \right\|_1
    =
    2\sqrt{
        \langle X_1,X_1\rangle_{\rho_0}^{\mathrm{S},(e)}
        \langle X_2,X_2\rangle_{\rho_0}^{\mathrm{S},(e)}
        -
        \left(\langle X_1,X_2\rangle_{\rho_0}^{\mathrm{S},(e)}\right)^2 
    }   =
    2\sqrt{\det\Phi[\mathsf X]} .
\end{equation}
Notice that this identity was already implicitly used to show the attainability result in Eq.~\eqref{eq:final_Nagaoka_optimalPOVM}.

Thus, the remaining optimisation is rewritten in terms of the Gram matrix and inner product as
\begin{equation}
    \min_{X_1,X_2} \tr \Phi[\mathsf{X}] + 2\sqrt{\det\Phi[\mathsf{X}]} - 2 \langle X_1, \Delta S_1 \rangle_{\rho_0}^{\mathrm{S},(e)} - 2 \langle X_2, \Delta S_2 \rangle_{\rho_0}^{\mathrm{S},(e)} . 
\end{equation}

Since \(\Phi[\mathsf X]\) is a positive semidefinite \(2\times2\) matrix, if its eigenvalues are \(\alpha_1,\alpha_2\geq0\), then
\begin{equation}
    \tr\Phi[\mathsf X]+2\sqrt{\det\Phi[\mathsf X]}
    =
    \alpha_1+\alpha_2+2\sqrt{\alpha_1\alpha_2}
    =
    \left(\sqrt{\alpha_1}+\sqrt{\alpha_2}\right)^2
    =
    \left(\tr\sqrt{\Phi[\mathsf X]}\right)^2.
\end{equation}

To perform the last minimisation, choose an arbitrary real basis of $\mathcal{V}$, i.e. three linearly independent zero-mean Hermitian operators $\{F_a\}_{a=1}^3$.
Then every $X_i\in\mathcal{V}$ and every $\Delta S_i\in\mathcal{V}$ can be represented uniquely by three-dimensional vectors
\begin{equation}
    X_i=\sum_{a=1}^3 x_{ai}F_a,
    \qquad
    \Delta S_i=\sum_{a=1}^3 d_{ai}F_a.
\end{equation}
Then we can introduce the matrices $X,D\in\mathbb R^{3\times2}$ with columns $x_i=(x_{1i},x_{2i},x_{3i})^T$ and $d_i=(d_{1i},d_{2i},d_{3i})^T$, respectively.
In this basis, the inner product $\langle\cdot,\cdot\rangle_{\rho_0}^{\mathrm{S},(e)}$ is represented by the real positive-definite matrix
\begin{equation}
    G_{\rho_0,ab} :=
    \langle F_a,F_b\rangle_{\rho_0}^{\mathrm{S},(e)} .
\end{equation}
Therefore we can express the Gram matrices of $X_1, X_2$ and $\Delta S_1, \Delta S_2$ as
\begin{equation}
    \Phi[X]=X^TG_{\rho_0}X,
    \qquad
    \mathcal{H}=D^TG_{\rho_0}D,
\end{equation}
and the mixed product term as
\begin{equation}
    \sum_{i=1}^2
    \langle X_i,\Delta S_i\rangle_{\rho_0}^{\mathrm{S},(e)}
    =
    \tr X^TG_{\rho_0}D.
\end{equation}
The remaining optimisation becomes
\begin{equation}
    \min_{X\in\mathbb R^{3\times2}}
    \left\{
        \left(\tr\sqrt{X^TG_{\rho_0}X}\right)^2
        -
        2\tr X^TG_{\rho_0}D
    \right\}.
    \label{eq:NH_remaining_matrix_form}
\end{equation}
Now we can introduce the matrices $Y:=G_{\rho_0}^{1/2}X$ and $ Z:=G_{\rho_0}^{1/2}D$ so that
\begin{equation}
    X^TG_{\rho_0}X=Y^TY,
    \qquad
    \tr X^TG_{\rho_0}D =\tr Y^TZ .
\end{equation}
Hence Eq.~\eqref{eq:NH_remaining_matrix_form} becomes
\begin{equation}
    \min_Y
    \left\{
        \|Y\|_1^2-2\tr Y^TZ
    \right\},
\end{equation}
where $\|Y\|_1:=\tr\sqrt{Y^TY}$ is the trace (nuclear) norm of a real matrix (i.e. the sum of the singular values).
By the standard duality between the trace norm and the operator norm, i.e. the maximum singular value of a matrix denoted as $\| \cdot \|_\infty$, we have:
\begin{equation}
    \tr Y^TZ \leq \|Y\|_1\,\|Z\|_{\infty},
\end{equation}
which implies
\begin{equation}
    \|Y\|_1^2-2\tr Y^TZ
    \geq
    \|Y\|_1^2-2\|Y\|_1\,\|Z\|_{\infty}
    \geq
    -\|Z\|_{\infty}^2 .
\end{equation}
where the last inequality follows by minimising the scalar $t^2-2t\|Z\|_\infty$ over $t\geq0$.
Equality is attained by taking \(Y=\|Z\|_\infty uv^T\), where \(u,v\) are left and right singular vectors associated with the largest singular value of \(Z\).
Consequently,
\begin{equation}
    \min_Y
    \left\{
        \|Y\|_1^2-2\tr Y^TZ
    \right\}
    =
    -\|Z\|_{\infty}^2 .
\end{equation}
Finally, by noticing that $Z^TZ = D^TG_{\rho_0}D = \mathcal{H}$, we have $\|Z\|_{\infty}^2 = \lambda_{\max}(\mathcal{H})$.
Thus the zero-mean part of the optimisation contributes
\(-\lambda_{\max}(\mathcal{H})\), and substituting into
Eq.~\eqref{eq:NHqubitexplicit-zeromean-objective} gives
\begin{equation}
    \mathcal L_{\mathrm{NH}}
    =
    \lambda-\mu_1^2-\mu_2^2-\lambda_{\max}(\mathcal{H}).
\end{equation}

The equivalent expression in Eq.~\eqref{eq:NH_qubit_closed_comm} comes from rewriting $\lambda_{\min}(\mathcal{H})$ in terms of $\tr \mathcal{H} = \lambda_{\min}(\mathcal{H}) + \lambda_{\max}(\mathcal{H}) $ and $\det \mathcal{H}  = \lambda_{\min}(\mathcal{H}) \lambda_{\max}(\mathcal{H})$, and using the identity in Eq.~\eqref{eq:Nagaoka-identity} to express the determinant in terms of the commutator.

\end{proof}

\section{Matrix lower bounds from monotone metrics}
\label{app:matrix-lower-bounds}

\subsection{Brief overview of monotone metrics}
\label{app:monotone-metrics}
In this subsection we follow the notation of Ref.~\cite[Ch.~6.1]{Hayashi2017c}, but see Ref.~\cite{Scandi2025} for a recent survey on the topic of monotone metrics.
For a given quantum state $\rho$, we define an inner product as follows:
\begin{equation}
\langle Y,\,X\rangle_{\rho}^{(e)} \;\coloneq\;\Tr\left[ Y^\dag \,E_{\rho}[X]\right],
\end{equation}
where $E_{\rho}$ is a linear map satisfying
\begin{align}
\Tr\left[ X^\dag E_\rho[X] \right]
&\ge 
0, \\
\Tr\left[ Y^\dag E_\rho[X] \right]
&= 
\Tr\left[ E_\rho[Y]^\dag\, X \right],
\end{align}
and
\begin{align}
E_\rho\left[U^\dag X U\right]
&= U^\dag\,E_{\,U\rho U^\dag}\!\left[X\right]\,U,\\
E_\rho\left[\id\right]
&= \rho,
% {\color{red} E_{\,\rho\otimes\rho'}\!\left[X\otimes X'\right]}
% &= {\color{red} E_\rho\left[X\right] \otimes E_{\rho'}\left[X'\right].
% \text{ [This is incorrect, even if written in Hayashi's book! (doesn't affect us)]}}
\end{align}
Following Hayashi's formulation, these maps and inner products are defined on arbitrary finite-dimensional operators, so $X$ and $Y$ need not be Hermitian.
The choice of the map $E_{\rho}$ determines the inner product.
We will primarily work with a generic one; however, three notable examples are:
\begin{align}
E_{\rho,\mathrm{S}}[X] & = \frac{1}{2}\left( X \rho + \rho X \right) \\ 
E_{\rho,\mathrm{R}}[X] & = X \rho \\ 
E_{\rho,\mathrm{sq}}[X] & = \sqrt{\rho} X \sqrt{\rho},
\end{align}
which we will call ``symmetric'', ``right'' and ``square-root'', respectively.
As in the main text, we assume $\rho \succ 0$ and for the standard monotone maps we will consider next it ensures that $E_\rho$ is invertible, so that we can introduce the dual inner product in terms of the inverse linear map $E_{\rho}^{-1}$ without further regularisations:
\begin{equation}
\langle A ,\, B \rangle_{\rho}^{(m)} \;=\;\Tr\left[ E_{\rho}^{-1}[A]^\dag B \right] ;
\end{equation}
it is denoted by the superscript $(m)$, following Hayashi's notation.
More specifically, in what follows we consider inner products with the standard
classical normalisation: when $\rho$ and $X,Y$ are diagonal in the same basis, the dual inner product reduces to the classical Fisher-type
expression $\langle X,Y\rangle_{\rho}^{(m)} = \sum_k \frac{X_{kk}^* Y_{kk}}{\rho_{kk}}$.

One of the crucial properties of such dual inner products is the monotonicity under the action of quantum channels, which makes them a good measure of information.
In the following, we restrict to choices of $E_\rho$ belonging to the standard class of monotone quantum metrics characterised by Petz~\cite{Petz1996a,Scandi2025}, or more generally to maps whose dual inner products satisfy the data-processing inequality\footnote{
Petz's work gives a complete characterisation of normalised monotone Riemannian metrics on the tangent space of full-rank quantum states,
i.e., on Hermitian traceless operators.
Here we also use the associated inner products on the full operator space, so we explicitly assume that the chosen extension continues to satisfy
Eq.~\eqref{eq:inner_monotonicity}.}:
\begin{equation}
  \label{eq:inner_monotonicity}
  \langle A, A \rangle_{\rho}^{(m)} \geq \langle \Lambda[A],  \Lambda[A] \rangle_{\Lambda[\rho]}^{(m)} \,,
\end{equation}
where $\Lambda$ is a quantum channel, i.e., a completely positive trace-preserving map, such that also the output state is full rank $\Lambda[ \rho] \succ 0$.
In the derivation below we will only use classical normalisation and this data-processing inequality. 

For normalised monotone metrics in the Petz class, another important property which holds for Hermitian operators $A=A^\dag$ is that the norm induced by this class of inner products is minimal and maximal for the symmetric and right versions, respectively:
\begin{equation}
  \label{eq:inner_hierarchy}
  \langle A , A \rangle_{\rho}^{\mathrm{S},(m)} \,  \leq \langle A , A \rangle^{(m)}_{\rho} \,  \leq \langle A , A \rangle_{\rho}^{\mathrm{R},(m)} \, .
\end{equation}
Eq.~\eqref{eq:sym_vs_sqrt_monotone_ineq} in the main text follows from the first inequality.
In the derivation of the matrix lower bounds below, where complex linear combinations of posterior moments are used, we only rely on the monotonicity property~\eqref{eq:inner_monotonicity}, which holds
for arbitrary operators.

\subsection{Posterior mean operators}
\label{app:posterior-mean-operators}

Introducing the MSL matrix, with elements
\begin{equation}
    \Sigma_{ij} = \int \! \mathrm{d} x \,  \int \mathrm{d} \params \, p(\params, x)\, ( \tilde{\theta}_{i}(x) - \theta_i ) ( \tilde{\theta}_{j}(x) - \theta_j ) \, ,
\end{equation}
where  $p(\params, x) = p(\boldsymbol{\theta}) \, p(x | \boldsymbol{\theta})$, the scalar MSL can be written as
\begin{equation}
    \mathcal{L} = L^{ij} \Sigma_{ji} = \tr L \Sigma \,.
\end{equation}

Classically, and thus also for quantum problems where the measurement is fixed, the optimal Bayes estimator that minimises the multiparameter MSL matrix is the PM:
\begin{equation}
    \tilde{\theta}_{\mathrm{opt},i}(x) = \int \! \mathrm{d}\params \, p( \params | x) \, \theta_i = \frac{ \int \! \mathrm{d}\params \, p(\params) \, p( x  | \params ) \, \theta_i}{ \int \! \mathrm{d} \params \, p(\params) \, p( x | \params ) },
\end{equation}
which was already defined in Eq.~\eqref{eq:classical_PM_estimator} of the main text.
Applying this estimator, the optimal MSL matrix has elements
\begin{equation}
    \Sigma_{ij} = \int \! \mathrm{d} \params \, p (\params) \, \theta_i \theta_j - \int \! \mathrm{d}x \,  \frac{ \left[ \int \! \mathrm{d}\params \, p(\params) \, p( x  |  \params ) \, \theta_i \right] \left[ \int \! \mathrm{d}\params \, p(\params) \, p( x  |  \params ) \, \theta_j \right] }{ \int \! \mathrm{d} \params \, p(\params) \, p( x | \params ) } \, .
\end{equation}

Following Ref.~\cite{rubio2020bayesian}, we now proceed to consider the positive scalar
% \begin{equation}
%     \boldsymbol{u}^T \Sigma \boldsymbol{u} = u^i u^j \Sigma_{ij}  =  \int \! \mathrm{d} \params \, p (\params) \, \theta_{\boldsymbol{u}}^2 - \int \! \mathrm{d}x \,  \frac{ \left[ \int \! \mathrm{d}\params \, p(\params) \, p( x  |  \params ) \, \theta_{\boldsymbol{u}} \right]^2 }{ \int \! \mathrm{d} \params \, p(\params) \, p( x | \params ) },
% \end{equation}
\begin{equation}
    \boldsymbol{u}^\dag \Sigma \boldsymbol{u} = u^{*i} u^j \Sigma_{ij}  =  \int \! \mathrm{d} \params \, p (\params) \, |\theta_{\boldsymbol{u}}|^2 - \int \! \mathrm{d}x \,  \frac{ \left| \int \! \mathrm{d}\params \, p(\params) \, p( x  |  \params ) \, \theta_{\boldsymbol{u}} \right|^2 }{ \int \! \mathrm{d} \params \, p(\params) \, p( x | \params ) },
\end{equation}  
where $\boldsymbol{u} \in \mathbb{C}^d$ is a complex vector and $\theta_{\boldsymbol{u}} = \boldsymbol{u}^\dag \params = u^{*i} \theta_i$ is a linear combination of parameters.
For quantum systems, the outcome probability for a given value of the parameters is obtained from Born's rule $p(x|\params) = \Tr \left[ M(x) \rho(\params)  \right]$, which leads to a complexified version of Eq.~(6) in Ref.~\cite{rubio2020bayesian}:
\begin{equation}
    \boldsymbol{u}^\dag \Sigma \boldsymbol{u}  = \int \! \mathrm{d} \params \, p (\params) \, |\theta_{\boldsymbol{u}}|^2 - \int \! \mathrm{d}x \,  \frac{ \left| \Tr \left[ M(x) \bar{\rho}_{\boldsymbol{u}} \right] \right|^2 }{\Tr \left[ M(x) \rho_0 \right] } \, ,
      \label{eq:eq6RubioMulti2020}
\end{equation}
where $\bar{\rho}_{\boldsymbol{u}} = u^{*i} \bar{\rho}_i = \int \! \mathrm{d} \params \, p(\params) \, \rho(\params) \, \theta_{\boldsymbol{u}}$.

The same result can be arrived at from the formalism of Sec.~\ref{sec:opt-quantum-strategies} with the posterior mean estimator Eq.~\eqref{eq:PMestimator}
being
\[
\tilde{\theta}_j(x) = \frac{\Tr[\bar{\rho}_j M(x)]}{\Tr[\rho_0 M(x)]},
\]
assuming a parameter-independent measurement.
As such, the operators \( M_j \) and \( M_{jk} \) introduced in Eq.~\eqref{eq:measurement-moments} take the form
\begin{equation}
\begin{aligned}
    M_i &= \int \mathrm{d}x\, M(x)\, \frac{\Tr[\bar{\rho}_i M(x)]}{\Tr[\rho_0 M(x)]} , &
    M_{ij} &= \int \mathrm{d}x\, M(x)\, \frac{\Tr[\bar{\rho}_i M(x)] \Tr[\bar{\rho}_j M(x)]}{\Tr[\rho_0 M(x)]^2};
\end{aligned}
\label{eq:measurement-moments-pme}
\end{equation}
the application in Eq.~\eqref{eq:msl_compact} then leads to~Eq.~\eqref{eq:eq6RubioMulti2020}.

A crucial observation from Ref.~\cite{rubio2020bayesian} is that the second term resembles the classical Fisher information and so it is possible to apply a Braunstein--Caves--like inequality~\cite{Braunstein1994} and obtain a matrix SPM lower bound on $\Sigma$.
Here, we take a different route and use a more general family of inequalities which stem from the monotonicity of the inner products defined in the previous sections.
This allows us to derive a whole family of matrix lower bounds, similarly to the case of local QET.

For simplicity, we focus on a discrete set of measurement outcomes $x$, which can always be encoded on a register composed by an orthonormal basis of vectors $\{\ket{e_x}\}$.
We introduce the measurement channel
\begin{equation}
\kappa[ \bullet ]  \coloneq \sum_x \Tr [ M(x) \bullet  ] | e_x \rangle \langle e_x | \, ;
\end{equation}
this is an entanglement-breaking quantum channel that formally represents the map between quantum states and probability distributions (diagonal quantum states) implemented by the POVM with elements $M(x)$.
Outcomes with $\Tr[M(x)\rho_0]=0$ can be discarded, or equivalently
one works on the support of $\kappa[\rho_0]$.
Applying this channel to $\rho_0$ and $\bar{\rho}_{\boldsymbol{u}}$ we obtain
\begin{equation}
\begin{split}
  \kappa[ \rho_0 ] &= \sum_x \Tr \left[ M(x) \rho_0 \right] |e_x \rangle \langle e_x |,  \\
  \kappa[ \bar{\rho}_{\boldsymbol{u}} ] &= \sum_x \Tr \left[ M(x) \bar{\rho}_{\boldsymbol{u}} \right] |e_x \rangle \langle e_x |  .
  \end{split}
\end{equation}
These operators are diagonal in the orthonormal basis $\{\ket{e_x}\}$ and they commute with each other; hence, by the classical normalisation stated above, the dual inner product gives
exactly the second term of Eq.~\eqref{eq:eq6RubioMulti2020}:
\begin{equation}
  \langle \kappa[ \bar{\rho}_{\boldsymbol{u}} ] , \kappa[ \bar{\rho}_{\boldsymbol{u}} ]  \rangle^{(m)}_{\kappa[ \rho_0 ] } = \Tr \left[ E_{\kappa[ \rho_0 ]}^{-1}(  \kappa[ \bar{\rho}_{\boldsymbol{u}} ]  )^\dag  \kappa[ \bar{\rho}_{\boldsymbol{u}} ]     \right] = \sum_{x} \frac{ \left| \Tr \left[ M(x) \bar{\rho}_{\boldsymbol{u}} \right] \right|^2 }{ \Tr \left[  M(x) \rho_0 \right]  }.
\end{equation}
Therefore, by the monotonicity property~\eqref{eq:inner_monotonicity} we finally obtain that
\begin{equation}
  \sum_{x} \frac{ \left| \Tr \left[ M(x) \bar{\rho}_{\boldsymbol{u}} \right] \right|^2 }{ \Tr \left[ M(x) \rho_0 \right]  }  \leq  \langle \bar{\rho}_{\boldsymbol{u}} , \bar{\rho}_{\boldsymbol{u}} \rangle^{(m)}_{\rho_0}  = \langle L_{\boldsymbol{u}} , L_{\boldsymbol{u}}  \rangle_{\rho_0}^{(e)},
  \label{eq:genericPMineq}
\end{equation}
where now $L_{\boldsymbol{u}}$ is a generic PM operator obtained as
\begin{equation}
  L_{\boldsymbol{u}} \coloneq E^{-1}_{\rho_0}[\bar{\rho}_{\boldsymbol{u}}] = u^{*i}  E^{-1}_{\rho_0}[ \bar{\rho}_i ] =  u^{*i} L_{i},
\end{equation}
where we have also introduced PM operators for each original parameter, satisfying the equations $E_{\rho_0}( L_i ) = \bar{\rho}_i$.
Such PM operators generalise the symmetric version introduced in the main text in terms of the map $E_{\rho , \mathrm{S}}$.
For example, they also include the operators introduced in Ref.~\cite[Theorem~8]{Zhang2024g}, where they are called ``$\lambda$ logarithmic derivatives''.

From the inequality in Eq.~\eqref{eq:genericPMineq}, the lower bound for the scalar introduced before immediately follows:
\begin{equation}
\boldsymbol{u}^\dag \Sigma \boldsymbol{u} \geq \int \! \mathrm{d} \params \, p (\params) \, \left| \theta_{\boldsymbol{u}} \right|^2 -  \boldsymbol{u}^\dag \mathcal{K} \boldsymbol{u},  
\end{equation}
where we have introduced the complex-valued Hermitian positive semidefinite matrix with elements
\begin{equation}
    \mathcal{K}_{ij} \coloneq  \langle \bar{\rho}_{j} , \bar{\rho}_{i} \rangle^{(m)}_{\rho_0}  = \langle L_{j} , L_{i}  \rangle_{\rho_0}^{(e)} = \Tr [ L_j^\dag E_{\rho_0}[ L_i ]  ] \,; 
\end{equation}
notice that with this index convention,
$\boldsymbol u^\dag\mathcal K\boldsymbol u
=\langle\bar\rho_{\boldsymbol u},
\bar\rho_{\boldsymbol u}\rangle_{\rho_0}^{(m)}\geq0$.
By choosing the map $E_{\rho , \mathrm{S}}$, this matrix becomes the real-valued SPM matrix introduced in Ref.~\cite{rubio2020bayesian}.
However, for other choices, such as $E_{\rho , \mathrm{R}}$, this matrix is still Hermitian but complex-valued.
Regardless of the particular choice of monotone metric, since the scalar inequality holds for an arbitrary choice of $\boldsymbol{u} \in \mathbb{C}^d$, we obtain the family of matrix lower bounds on the MSL matrix
\begin{equation}
    \label{eq:matrix_ineq_MSE}
    \Sigma \succeq \int \mathrm{d}\params\, p(\params)\, \params \params^T  - \mathcal{K} \;.
\end{equation}
This is again a generalisation of the SPM matrix bound obtained in Ref.~\cite{rubio2020bayesian}.

\subsection{Scalar bounds from matrix inequalities}

We start from the matrix inequality in Eq.~\eqref{eq:matrix_ineq_MSE} and denote the right-hand side by the symbol $\mathcal{B} \coloneq \int \mathrm{d}\params\, p(\params)\, \params \params^\top - \mathcal{K}$,
where $\mathcal{K} \geq 0 $ is a \emph{complex-valued} Hermitian positive semidefinite matrix.
We now want to show that also the imaginary part of $\mathcal{K}$ sets constraints on the minimum MSL $\mathcal{L}$.
This fact is well-known~\cite{holevo2011probabilistic}; for completeness, we provide a proof here, roughly following Ref.~\cite{demkowicz2020multi}.

First, let us rewrite the function to minimise, subject to the constraint \eqref{eq:matrix_ineq_MSE}, as the trace of a symmetric matrix $\tr  L \Sigma  = \tr \sqrt{L} \Sigma \sqrt{L}$.
From~\eqref{eq:matrix_ineq_MSE} we can obtain the two matrix inequalities
\begin{equation}
    \sqrt{L} \, \Sigma \sqrt{L} \geq \sqrt{L} \left( \Re{\mathcal{B}} \pm i \Im{\mathcal{B}} \right) \sqrt{L}, 
\end{equation}
where the two signs correspond to taking the transpose of the two sides of the original equation.
Here $\Re\mathcal B$ and $\Im\mathcal B$ denote the entrywise real
and imaginary parts, so that $\Re\mathcal B$ is real symmetric and
$\Im\mathcal B$ is real antisymmetric.
Now, this matrix inequality implies
\begin{equation}
    \sum_k \boldsymbol{z}_k^\dag   \sqrt{L} \left( \Sigma - \Re{\mathcal{B}} \right) \sqrt{L} \boldsymbol{z}_k \geq  \pm \sum_k \boldsymbol{z}_k^\dag   \left( i \sqrt{L} \Im{\mathcal{B}} \sqrt{L} \right)  \boldsymbol{z}_k \, ,
\end{equation}
for any set of complex-valued vectors $\{ \boldsymbol{z}_i \}$.
Since $i \sqrt{L} \Im{\mathcal{B}} \sqrt{L}$ is Hermitian, we can choose its eigenbasis as the vectors  $\boldsymbol{z}_i$; noting the corresponding eigenvalues as $\lambda_i$, we obtain
\begin{align}
    \tr  L ( \Sigma - \Re{\mathcal{B}} )  \geq \sum_i | \lambda_i | = \tr   \vert \sqrt{L} \Im{\mathcal{B}} \sqrt{L}  \vert   =  \Vert \sqrt{L} \Im{\mathcal{B}} \sqrt{L} \Vert_1 .
\end{align}
The absolute values are obtained by applying the positive inequality
separately to eigenvectors with positive eigenvalues and the negative
inequality separately to eigenvectors with negative eigenvalues,
and then summing the resulting inequalities.
The matrix $\vert \sqrt{L}\Im{\mathcal{B}}\sqrt{L}\vert$, where $\vert A\vert\coloneq\sqrt{A^\dag A}$ denotes the operator absolute value, is a positive semidefinite Hermitian matrix that can be diagonalised in the same eigenbasis as $i\sqrt{L}\Im{\mathcal{B}}\sqrt{L}$, with eigenvalues $|\lambda_i|$.
Moreover, $\Vert A\Vert_1=\tr \vert A\vert
=\tr \sqrt{A^\dag A} $ is the Schatten 1-norm, also known as the trace norm or nuclear norm.
This inequality for the trace of the real part of a positive semidefinite complex matrix is known as the Belavkin--Grishanin inequality~\cite{Belavkin1973,Tsang2019}.

Assuming $L \succ 0$ (otherwise one should restrict to the support of $L$), the minimum can be achieved by setting 
\begin{equation}
    \label{eq:sigma_opt}
    \Sigma = \Re{\mathcal{B}} + \sqrt{L^{-1}} \vert \sqrt{L} \Im{\mathcal{B}} \sqrt{L}  \vert \sqrt{L^{-1}} \, , 
\end{equation}
as seen from direct computation.

In conclusion, we have proven that 
\begin{equation} 
    \min_{ \Sigma \geq \mathcal{B} } \tr  L \Sigma = \tr L \Re{\mathcal{B}}  +
    \Vert \sqrt{L} \Im{\mathcal{B}} \sqrt{L} \Vert_1 .
\end{equation}
Substituting back the original right-hand side of the inequality~\eqref{eq:matrix_ineq_MSE}, we get
\begin{equation}
    \mathcal{L} \geq \lambda - \Tr[ L \Re{\mathcal{K}} ] + \Vert \sqrt{L} \Im{\mathcal{K}} \sqrt{L} \Vert_1,
\end{equation} 
since $\lambda = \tr  L \int \mathrm{d}\params \, p(\params) \, \params \params^T = \int \mathrm{d}\params \, p(\params) \, \params^T L\, \params $, and $\Im{\mathcal{B}}= - \Im{\mathcal{K}}$, given that the other matrix terms are real-valued by assumption.
In general, this bound is not necessarily tight, as it is not always possible to find a measurement and estimator that produce a MSL matrix as in \eqref{eq:sigma_opt}.

We finally note that the bound in Ref.~\cite[Theorem~8]{Zhang2024g} in terms of the $\lambda$ logarithmic derivatives is also in the class of scalar bounds derived in this Appendix.

\section{Phase-dephasing multi-copy expressions}%
\label{app:phase-dephasing_expressions}
For the phase-dephasing model introduced in Sec.~\ref{sec:phase-dephasing}, the full single-copy zeroth and first state moments are
\begin{subequations}
\begin{align}\label{eq:state-momenta-phase-dephasing}
    \rho_0 &=
    \frac{1}{2}\left\lbrace\mathbb{I}_2 + \frac{\sin\left(\frac{W_1}{2}\right) \left[\Ei(-W_2) - \Ei\left(-\frac{1}{W_2}\right)\right]}{W_1 \log(W_2)} \sigma_x\right\rbrace,
    \\
    \bar{\rho}_1 &=
    \frac{1}{2 W_1 \log(W_2)} \left[\frac{W_1}{2} \cos \left(\frac{W_1}{2}\right) - \sin\left(\frac{W_1}{2}\right)\right]
    \left[\Ei \left(-W_2\right) - \Ei \left(-\frac{1}{W_2}\right)\right] \sigma_y,
    \\
    \bar{\rho}_2 &=
    \frac{\sin(W_1/2)}{2W_1}
    \left(
    \frac{
    W_2 \Xi(-W_2)
    -\frac{1}{W_2} \Xi(-\frac{1}{W_2}) 
    }{2\log(W_2)}
    +
    \frac{\Ei(-W_2)+\Ei(-\frac{1}{W_2})}{2} - \gamma
    \right)
    \sigma_x,
\end{align}
\end{subequations}
where $\Ei(z)$ denotes the exponential integral, 
\(
    \Xi(z) = \displaystyle\sideset{_3}{_3}{\mathop{F}}((1,1,1),(2,2,2),z)
\)
is a generalised hypergeometric function, and $\gamma$ is Euler's constant.

By virtue of the tensor product structure---although the moments themselves are not direct tensor products---the multi-copy state \( \rho(\params) = \varrho(\params)^{\otimes n} \) consists of terms
\(
    \prod\limits_{j,k=1}^d [\varrho_{j,k}(\params)]^{\kappa_{j,k}}
\)
for a \( d \)-dimensional system.
For the multi-copy phase and dephasing example of Sec.~\ref{sec:phase-dephasing}, the pertinent integrals for the respective moments are given below.
For \( \rho_0 \):
\begin{subequations}
\begin{align}
\int \mathrm{d}\params\, 
\frac{
[\varrho_{1,1}(\params)]^{j}
[\varrho_{2,2}(\params)]^{n}
}{\theta_2 W_1 \log W_2}
&=
\frac{1}{2^{j+n}},
\label{eq:pdp:rho0:km0} \\
\int \mathrm{d}\params\, 
\frac{
[\varrho_{1,1}(\params)]^{j}
[\varrho_{1,2}(\params)]^{k}
[\varrho_{2,1}(\params)]^{k}
[\varrho_{2,2}(\params)]^{n}
}{\theta_2 W_1 \log W_2}
&=
\frac{
\Ei(-2kW_2) - \Ei\left(-\frac{2k}{W_2}\right)
}{2^{j+2k+n+1} \log W_2},
\label{eq:pdp:rho0:keqm} \\
\int \mathrm{d}\params\, 
\frac{
[\varrho_{1,1}(\params)]^{j}
[\varrho_{1,2}(\params)]^{k}
[\varrho_{2,1}(\params)]^{m}
[\varrho_{2,2}(\params)]^{n}
}{\theta_2 W_1 \log W_2}
&=
\frac{
\left[
\Ei(-(k+m)W_2) - \Ei\left(-\frac{(k+m)}{W_2}\right)
\right]
\sin\left( \frac{(k-m)W_1}{2} \right)
}{2^{j+k+m+n} (k-m) W_1 \log W_2},
\label{eq:pdp:rho0:knem}
\end{align}
\end{subequations}
where Eq.~\eqref{eq:pdp:rho0:keqm} holds for \( k>0 \), and Eq.~\eqref{eq:pdp:rho0:knem} for \( k \neq m \).
For \( \bar{\rho}_1 \):
\begin{subequations}
\begin{align}
\int \mathrm{d}\params\, 
\frac{
[\varrho_{1,1}(\params)]^{j}
[\varrho_{1,2}(\params)]^{k}
[\varrho_{2,1}(\params)]^{k}
[\varrho_{2,2}(\params)]^{n}
\theta_1
}{\theta_2 W_1 \log W_2}
&=
0,
\label{eq:pdp:rho1:keqm} \\
\int \mathrm{d}\params\, 
\frac{
[\varrho_{1,1}(\params)]^{j}
[\varrho_{1,2}(\params)]^{k}
[\varrho_{2,1}(\params)]^{m}
[\varrho_{2,2}(\params)]^{n}
\theta_1
}{\theta_2 W_1 \log W_2} 
 &= 
\frac{
i\left[
\Ei(-(k+m)W_2) - \Ei\left(-\frac{k+m}{W_2}\right)
\right]
}{
2^{j+k+m+n} (k-m)^2 W_1 \log W_2 }
\label{eq:pdp:rho1:knem}
\\ \nonumber
&\mkern32mu\times\left[
\sin\left( \frac{(k-m)W_1}{2} \right)
- \frac{(k-m)W_1}{2}
\cos\left( \frac{(k-m)W_1}{2} \right)
\right]
,
\end{align}
\end{subequations}
where Eq.~\eqref{eq:pdp:rho1:knem} holds for \( k \neq m \).
For \( \bar{\rho}_2 \):
\begin{subequations}
\begin{align}
\int \mathrm{d}\params\, 
\frac{
[\varrho_{1,1}(\params)]^{j}
[\varrho_{2,2}(\params)]^{n}
\log\theta_2
}{\theta_2 W_1 \log W_2}
&=
0,
\label{eq:pdp:rho2:km0} \\
\int \mathrm{d}\params\, 
\frac{
[\varrho_{1,1}(\params)]^{j}
[\varrho_{1,2}(\params)]^{k}
[\varrho_{2,1}(\params)]^{k}
[\varrho_{2,2}(\params)]^{n}
\log\theta_2
}{\theta_2 W_1 \log W_2}
&=
\frac{
\xi_{2k}(W_1,W_2)
}{2^{j+2k+n+1}} ,
\label{eq:pdp:rho2:keqm}
\\
\int \mathrm{d}\params\, 
\frac{
[\varrho_{1,1}(\params)]^{j}
[\varrho_{1,2}(\params)]^{k}
[\varrho_{2,1}(\params)]^{m}
[\varrho_{2,2}(\params)]^{n}
\log\theta_2
}{\theta_2 W_1 \log W_2}
&=
\frac{ 
\sin\left(\frac{(k-m)W_1}{2} \right) \xi_{k+m}(W_1,W_2)
}{ 2^{j+k+m+n} W_1 (k-m) }
\label{eq:pdp:rho2:knem}
\end{align}
\end{subequations}
where
\[
\xi_{q}(W_1,W_2)=
\frac{q}{2\log W_2}
\left(
W_2 \Xi(-q W_2)
-\frac{\Xi\left(-\frac{q}{W_2}\right)}{W_2}
\right)
+
\frac{
\Ei \left(-q W_2\right) + \Ei \left(-\frac{q}{W_2}\right)
}{2}
-\log(q) - \gamma,
\]
and Eq.~\eqref{eq:pdp:rho2:keqm} holds for \( k>0 \), and Eq.~\eqref{eq:pdp:rho2:knem} for \( k \neq m \).
\end{document}